\documentclass{aa}
\usepackage{amssymb,amsmath,graphicx,xspace,natbib}
\usepackage[varg]{txfonts}
\bibpunct{(}{)}{;}{a}{}{,} 
\newcommand{\eg}{{e.g.\/}}
\newcommand{\ie}{{i.e.\/}}

\newcommand{\cf}{{c.f.\/}}
\newcommand{\etc}{{etc.\/}}
\newcommand{\expon}[1]{\ensuremath{\times 10^{#1}}}
\newcommand{\azeus}{\textsf{AZEuS}\xspace}
\newcommand{\zeus}{\textsl{ZEUS}\xspace}

\newcommand{\del}{\ensuremath{\partial}}
\newcommand{\petsc}{\textsf{PETSc}\xspace}
\newcommand{\er}{\ensuremath{E_\mathrm{R}}}
\newcommand{\fr}{\ensuremath{\vec{F}_\mathrm{R}}}
\newcommand{\etot}{\ensuremath{E_\mathrm{T}}}
\newcommand{\kappae}{\ensuremath{\kappa_\mathrm{E}}}
\newcommand{\kappap}{\ensuremath{\kappa_\mathrm{P}}}
\newcommand{\kappar}{\ensuremath{\kappa_\mathrm{R}}}
\newcommand{\arad}{\ensuremath{a_\mathrm{R}}}
\usepackage[%
pdfauthor={Ramsey \& Dullemond},
pdftitle={Ramsey \& Dullemond: RHD with AZEuS},
pdfstartview=FitH,
linkcolor=blue,
anchorcolor=blue,
citecolor=blue,
filecolor=blue,
menucolor=blue,
urlcolor=blue,
colorlinks=true]{hyperref}
\begin{document}
\title{Radiation hydrodynamics including irradiation and \mbox{adaptive mesh refinement} with \azeus}
\subtitle{I. Methods}
\titlerunning{RHD with \azeus}
\author{J.~P.~Ramsey \and C.~P.~Dullemond}
\authorrunning{Ramsey \& Dullemond}
\institute{Universit\"at Heidelberg, Zentrum f\"ur Astronomie, Institut f\"ur Theoretische Astrophysik, Albert-\"Uberle-Str.\ 2, D-69120 Heidelberg, Germany\\ \email{ramsey@uni-heidelberg.de}}
\date{\today}
%
%
\abstract
{}
{The importance of radiation to the physical structure of protoplanetary disks cannot be understated.  However, protoplanetary disks evolve with time, and so to understand disk evolution and by association, disk structure, one should solve the combined and time-dependent equations of radiation hydrodynamics.}
{We implement a new implicit radiation solver in the \azeus adaptive mesh refinement magnetohydrodynamics fluid code.  Based on a hybrid approach that combines frequency-dependent ray-tracing for stellar irradiation with non-equilibrium flux limited diffusion, we solve the equations of radiation hydrodynamics while preserving the directionality of the stellar irradiation.  The implementation permits simulations in Cartesian, cylindrical, and spherical coordinates, on both uniform and adaptive grids.} 
{We present several hydrostatic and hydrodynamic radiation tests which validate our implementation on uniform and adaptive grids as appropriate, including benchmarks specifically designed for protoplanetary disks.  Our results demonstrate that the combination of a hybrid radiation algorithm with \azeus is an effective tool for radiation hydrodynamics studies, and produces results which are competitive with other astrophysical radiation hydrodynamics codes.}
{}
\maketitle
\section{Introduction}
\label{sec:intro}

From Poynting-Robertson drag on dust grains to cosmological reionisation, and at every scale in between, there is no denying the importance of radiation in astrophysics.  In accretion disks around young stars in particular, the effects of radiative cooling and stellar irradiation are crucial to determining the thermal and physical structure of the disk (\eg, flaring; \citealt{chianggoldreich97}).

Until recently, models of these protoplanetary accretion disks that include radiation have been restricted to 1+1-D dynamic or 2- and 3-D static models (\eg, \citealt{gortietal09,brudereretal10,minetal11}).  However, many phenomena, such as planetary migration (\eg, \citealt{kleyetal09}), or the magneto-rotational instability (MRI; \eg, \citealt{flaigetal10}) are certainly dynamic and multidimensional, and so require time-dependent radiative transfer (RT) coupled to dynamics.

One approach is to self-consistently couple the equations of hydrodynamics (HD) and radiation transport.  Although the combined subject of radiation hydrodynamics (RHD) is treated with great detail in many textbooks (\eg, \citealt{mm84}), the development of practical methods for numerical modelling still remains an active area of astrophysical research.

A recent approach to RHD in accretion disks, made popular by \citet{kuiperetal10}, is to combine flux-limited diffusion (FLD) with a simple and fast ray-tracing approach for the direct irradiation by stellar photons.  In this algorithm, the direct stellar flux is followed until it is absorbed, and the resultant heat is added to the fluid.  Any subsequent re-emission is handled as diffuse radiation using FLD.  Alternatively, one could state that the stellar photons are followed until the surface of first absorption, whereupon they heat the fluid.  Treating the stellar flux in this hybrid manner preserves the directional properties of the stellar radiation significantly better than with FLD alone, and the result can even be competitive with more complicated and expensive Monte Carlo methods \citep{kuiperetal10, kuiperklessen13}.  It is also well-known that FLD is incapable of casting shadows \citep{hayesnorman03}, a potentially important effect for irradiated protoplanetary disks with puffed up inner rims (\eg, \citealt{dullemondetal01}), but which is possible with the hybrid irradiation + FLD algorithm.

Although this frequency-dependent algorithm is straightforward to implement and fast, the trade-off is that rays cannot travel in arbitrary directions: the irradiation flux is restricted to travelling parallel to coordinate axes.  This naturally presents challenges for general circumstances, in particular scattering of radiation or multiple sources, but when applied to accretion disks using spherical coordinates, it is very efficient.

In this regard, there are certainly more accurate methods for radiation hydrodynamics available, including the variable Eddington tensor factor (\eg, \citealt{jiangetal12}), M1 closure (\eg, \citealt{gonzalezetal07}), and Monte Carlo plus hydrodynamics methods (\eg, \citealt{haworthharries12}).  However, these methods are normally much more computationally expensive than the hybrid method discussed here.

It is also worth noting that the splitting of radiation into multiple components (whether diffusion, ray-tracing, stellar heating, or other) is not a new idea, and several examples in different contexts can be found in the literature (\eg, \citealt{wolfirecassinelli86,murrayetal94, edgarclarke03,whalennorman06, boleyetal07,dobbsdixonetal10}).

In this paper, we present a new radiation module for the \azeus adaptive mesh refinement (AMR) magnetohydrodynamics (MHD) code \citep{rcm12}, with a focus on protoplanetary disk (PPD) applications.  We have combined frequency-dependent ray-tracing \citep{kuiperetal10} with non-equilibrium (two-temperature) FLD (\eg, \citealt{bitschetal13,kolbetal13, flocketal13}) and, in addition, coupled it to AMR \citep{bergercolella89}.

While there are other AMR FLD codes available in the astrophysical literature\footnote{Nested-grid, or \emph{static} mesh refinement FLD codes are also available, \eg, \citet{tomidaetal13}.}, including, for \eg, ORION \citep{krumholzetal07}, CRASH \citep{vanderholst11}, CASTRO \citep{zhangetal11}, RAMSES \citep{commerconetal14}, and FLASH \citep{klassenetal14}, \azeus is the only AMR fluid code currently available which employs a \emph{fully} staggered mesh, where the momentum and magnetic field are both face-centred quantities.  Furthermore, as it is based on the \zeus family of codes, it derives from one of the best documented, tested, and widely used codes in astrophysics.  


The paper layout is as follows.  In \S \ref{sec:radhydro} we describe our algorithms for radiation hydrodynamics in the uniform grid case.  In \S \ref{sec:amr+rhd}, we present out extensions to both FLD and irradiation for use with AMR.  In \S \ref{sec:tests}, we perform tests to validate our algorithms.  Finally, in \S \ref{sec:discuss}, we summarise our results and present an outlook for future work.
\section{Radiation hydrodynamics in \azeus}
\label{sec:radhydro}
\azeus solves the time-dependent, frequency-integrated equations of RHD in the co-moving frame, to order ${\cal O}(v/c)$, and assuming local thermodynamic equilibrium:

\begin{align}
  \label{eq:continuity}
  \frac{\del \rho}{\del t} + \nabla\cdot\left(\rho\vec{v}\right) =&~ 0;\\
  \label{eq:momentum}
  \frac{\del\vec{s}}{\del t} + \nabla\cdot\left(\vec{s} \vec{v}\right) =&~ \nabla p - \nabla\cdot\tens{Q} - \rho\nabla\Phi + \nabla\cdot\tens{T} + \frac{\rho\kappa_\mathrm{F}}{c}\vec{F}_\mathrm{R};\\
  \label{eq:intenergy}
  \begin{split}
  \frac{\del e}{\del t} + \nabla\cdot\left(e\vec{v}\right) =& -p\nabla\cdot\vec{v} + c\rho\left(\kappa_\mathrm{E} \er - \kappa_\mathrm{P} \arad T^4\right)\\
  & + \tens{T}\!:\!\nabla\vec{v} - \tens{Q}\!:\!\nabla\vec{v};
  \end{split}\\
    \label{eq:radenergy}
  \frac{\del \er}{\del t} + \nabla\cdot\left(\er \vec{v}\right) =& -c\rho\left(\kappa_\mathrm{E} \er - \kappa_\mathrm{P} \arad T^4\right) - \nabla\cdot\vec{F}_\mathrm{R} - \tens{P}\!:\!\nabla\vec{v} ;\\
  \label{eq:totenergy}
  \begin{split}
  \frac{\del \etot}{\del t} + \nabla\cdot\Big( (\etot + &p)\vec{v} + \tens{Q}\cdot\vec{v} \Big) =  - \nabla\cdot\vec{F}_\mathrm{R} - \tens{P}\!:\!\nabla\vec{v} \\
  &\qquad\qquad\quad~~~ + \nabla\cdot\left(\tens{T}\cdot\vec{v}\right) + \frac{\rho\kappa_\mathrm{F}}{c}\vec{v}\cdot\vec{F}_\mathrm{R},
  \end{split}
\end{align}

where $\rho$ is the mass density, $\vec{v}$ is the velocity, $\vec{s} = \rho\vec{v}$ is the momentum density, $p$ is the thermal pressure, $\mathsf{Q}$ is the artificial viscous stress tensor \citep{vonneumannrichtmyer1950}, $\Phi$ is the gravitational potential, $\tens{T} = \rho\nu(\nabla\vec{v} + \nabla\vec{v}^{T} - \frac{2}{3}\tens{I}\nabla\cdot\vec{v})$ is the viscous stress tensor, $\nu$ is the kinematic viscosity, $\tens{I}$ is the identity matrix, $e$ is the internal energy density, $\er$ is the radiation energy density, $\etot = e + \frac{1}{2}\rho v^2 + \er$ is the total energy density, $T$ is gas temperature, $\vec{F}_\mathrm{R}$ is the radiation flux, $\tens{P}$ is the radiation pressure tensor, and $\kappap$, $\kappae$, $\kappa_\mathrm{F}$ are the Planck, radiation energy, and radiation flux mean opacities.  The radiation constant and the speed of light are $a_\mathrm{R}$ and $c$, respectively.  Although different values for $\kappap$ and $\kappae$ are permitted in the algorithm, all the tests presented here assume that the radiation behaves like a blackbody, and therefore $\kappap = \kappae$.

The first closure relation applied to the above equation set is the ideal gas law: $p = \left(\gamma - 1\right)e = \rho C_V T$, where $\gamma$ is the ratio of specific heats, $C_V = k / (\gamma - 1)\mu m_\mathrm{H}$ is the specific heat capacity at constant volume, $k$ is Boltzmann's constant, $\mu$ is the mean molecular weight, and $m_\mathrm{H}$ is the mass of hydrogen.

MHD is, of course, also available in \azeus, but we restrict the discussion herein to RHD for simplicity.  \azeus also includes the option to solve either the total energy equation or the internal energy equation.  It should be noted, however, that the radiation solver uses the internal energy density.  Furthermore, total energy will not generally be conserved in the co-moving frame.  For details on the MHD and AMR algorithms in \azeus, including test problems, we refer the reader to \citet{clarke96, clarke10, rcm12}, and references therein.
\subsection{Flux-limited diffusion}
\label{sub:fld}
In order to close equation set (\ref{eq:continuity})--(\ref{eq:radenergy}), we adopt the FLD approximation, where the radiative flux is replaced by Fick's law:
\begin{equation}
  \label{eq:fick}
  \vec{F}_\mathrm{R} = -\lambda(R)\frac{c}{\rho \kappa_\mathrm{R}}\nabla \er = -D\nabla \er;
\end{equation}
$\lambda(R)$ is the so-called flux limiter, $\kappar$ is the Rosseland mean opacity, and $D$ is the diffusion coefficient.  We assume that $\kappar = \kappa_\mathrm{F}$, a common approximation in the diffusion regime, and which preserves the correct energy and momentum transport.

By omitting a conservation equation for the radiative flux, and under the Eddington approximation ($\lambda = 1/3$), it is possible with Eq.\ (\ref{eq:fick}) to generate an unphysically high flux ($> c\er$), leading to an energy propagation speed that exceeds the speed of light.  Flux limiters, by design, correct for this.  One commonly used form for the flux limiter is given by \citep{lp81}:
\begin{equation}
  \label{eq:fluxlimiter}
  \lambda(R) = \frac{2 + R}{6 + 3R + R^2},
\end{equation}
where
\begin{equation}
  \label{eq:rvalue}
  R = \frac{|\nabla \er|}{\rho\kappar \langle\er\rangle},
\end{equation}
and $\langle\phantom{.}\rangle$ denotes a spatially-averaged quantity.  In Eq.\ (\ref{eq:fluxlimiter}), as $R$ becomes large, $\lambda \rightarrow 1/R$, and the flux is limited to $\leq c\er$.  Conversely, for small $R$, the flux limiter reduces to the Eddington approximation ($\lambda = 1/3$).  Flux-limiters currently implemented in \azeus include \citet{lp81}, \citet{minerbo78} and \citet{kley1989}.

Note that, in Eqs.\ (\ref{eq:fick}) and (\ref{eq:rvalue}), the radiative flux $\vec{F}_\mathrm{R}$, the diffusion coefficients $D$, and the quantity $R$ are all face-centred quantities while the product $\rho \kappar = \sigma_\mathrm{R}$ is intrinsically zone-centred.  Although, in principle, we could arithmetically average $\sigma_\mathrm{R}$ to the face, we instead adopt the surface formula of \citet{howellgreenough03} to improve results at sharp interfaces in optical depth.  For example, in the 1-direction:
\begin{equation}
  \label{eq:surfaceform}
  \sigma_{\mathrm{R},\, i+1/2} = \min \left[ \frac{\sigma_{\mathrm{R},i} + \sigma_{\mathrm{R},i+1}}{2}, \max \left(\frac{2\sigma_{\mathrm{R},i}\,\sigma_{\mathrm{R},i+1}}{\sigma_{\mathrm{R},i} + \sigma_{\mathrm{R},i+1}}, \frac{4}{3\Delta x_1} \right)\right].
\end{equation}

The radiation pressure tensor is assumed to have the form:
\begin{equation}
  \label{eq:radstress}
  \tens{P} = \tens{f} \er,
\end{equation}
where \tens{f} is called the Eddington tensor, and is given by \citep{turnerstone2001,hayesetal06}:
\begin{align}
  \label{eq:tenseddingfact}
  \tens{f} =&~ \frac{1}{2}(1 - f)\tens{I} + \frac{1}{2}(3f - 1)\hat{\vec{n}}\hat{\vec{n}};\\
  \label{eq:scaleddingfact}
  f =&~ \lambda + \lambda^2 R^2,
\end{align}
and $\hat{\vec{n}} = \nabla\er / |\nabla\er|$.

In \azeus, the coupled equations of radiation hydrodynamics are solved through the use of operator splitting, where the hydrodynamical terms are calculated explicitly, but the radiation terms are solved implicitly.  The coupled gas-radiation system of equations that we wish to solve is given by:
\begin{align}
  \label{eq:e_impsolv}
  \dfrac{\del e}{\del t} & = c\rho\left(\kappae \er - \kappap \arad T^4\right);\\
  \label{eq:er_impsolv}
  \dfrac{\del \er}{\del t} & = -c\rho\left(\kappae \er - \kappap \arad T^4\right) + \nabla\cdot\left( D\nabla \er \right) - \tens{P}\!:\!\nabla\vec{v},
\end{align}
and the advection, gravitational, compressive heating, and physical \& artificial viscous terms which appear in equations (\ref{eq:continuity})--(\ref{eq:radenergy}) are solved in the explicit part of the \azeus algorithm.

Using the ideal gas law, we can substitute the temperature for the internal energy density to find:
\begin{equation}
  \label{eq:temp_impsolv}
  \frac{\del T}{\del t} = \frac{c}{C_V}\left(\kappa_\mathrm{E} \er - \kappa_\mathrm{P} \arad T^4\right).
\end{equation}

We now write Eqs.\ (\ref{eq:er_impsolv}) and (\ref{eq:temp_impsolv}) in difference form:
\begin{align}
  \label{eq:temp_differenced}
  \frac{T^{n+1} - T^{n}}{\Delta t} &= \frac{c}{C_V} \left[ \kappae^{n+1} \er^{n+1} - \kappap^{n+1} \arad (T^{n+1})^4 \right];\\
  \label{eq:er_differenced}
  \begin{split}
  \frac{\er^{n+1} - \er^n}{\Delta t} &= -c\rho^n \left[ \kappae^{n+1} \er^{n+1} - \kappap^{n+1} \arad (T^{n+1})^4 \right]\\
  & + \nabla\cdot\left(D^n \nabla\er^{n+1}\right) - (\tens{f}^{\,n}\er^{n+1})\!:\!\nabla\vec{v}^n;
  \end{split}\\
  \label{eq:radflux_differenced}
  \begin{split}
  \nabla\cdot\left(D^n \nabla\er^{n+1}\right) &= \frac{1}{\Delta x_1}\left[ D^n_{i+1}\frac{E_{\mathrm{R},i+1}^{n+1} - E_{\mathrm{R},i}^{n+1}}{\Delta x_1} - D^n_{i}\frac{E_{\mathrm{R},i}^{n+1} - E_{\mathrm{R},i-1}^{n+1}}{\Delta x_1} \right]\\
  + \frac{1}{\Delta x_2}&\left[ D^n_{j+1}\frac{E_{\mathrm{R},j+1}^{n+1} - E_{\mathrm{R},j}^{n+1}}{\Delta x_2} - D^n_{j}\frac{E_{\mathrm{R},j}^{n+1} - E_{\mathrm{R},j-1}^{n+1}}{\Delta x_2} \right]\\
  + \frac{1}{\Delta x_3}&\left[ D^n_{k+1}\frac{E_{\mathrm{R},k+1}^{n+1} - E_{\mathrm{R},k}^{n+1}}{\Delta x_3} - D^n_{k}\frac{E_{\mathrm{R},k}^{n+1} - E_{\mathrm{R},k-1}^{n+1}}{\Delta x_3} \right]
  \end{split},
\end{align}
where $n$ and $n+1$ denote original and updated values within the implicit solve.  We have assumed Cartesian coordinates here for brevity ($x_1, x_2, x_3 = x, y, z$); the curvilinear factors associated with cylindrical and spherical coordinates are presented in Appendix A of \citet{rcm12}.  Also note that the diffusion coefficients (and therefore the flux limiter) and Eddington tensor $\tens{f}$ are time-lagged, which is a voluntary choice for stability and efficiency over accuracy.

As given, Eqs.\ (\ref{eq:temp_differenced}) - (\ref{eq:radflux_differenced}) form a non-linear system of equations.  If we now linearise the term in $(T^{n+1})^4$ by taking a Taylor expansion about $T^n$ \citep{commerconetal11}:
\begin{equation}
  \label{eq:expandt_np1}
  (T^{n+1})^4 \approx 4(T^n)^3 T^{n+1} - 3(T^n)^4,
\end{equation}

then $(T^{n+1})^4$ can be calculated in terms of known quantities only:
\begin{equation}
  \label{eq:t_np1}
  T^{n+1} = \frac{T^n + 3 a_\mathrm{R} c \frac{\Delta t}{C_V} \kappa_\mathrm{P}^n (T^n)^4 + \frac{\Delta t}{C_V} c \kappae^n E_\mathrm{R}^{n+1} }{1 + 4 \arad c \frac{\Delta t}{C_V} \kappap^n (T^n)^3},
\end{equation}
and thus eliminating the internal energy/temperature equation from the radiation system of equations.\footnote{An alternative approach would be to eliminate the internal energy equation via a partial LU decomposition (\eg, \citealt{hayesetal06,tomidaetal13}).}  Note that substituting Eqs.\ (\ref{eq:expandt_np1}) and (\ref{eq:t_np1}) into Eq.\ (\ref{eq:er_differenced}) would result in a linear system of equations (\eg, \citealt{commerconetal11}) if it were not for the fact that we account for the generally non-linear temperature-dependence of the mean opacities, $\kappap^{n+1}$ and $\kappae^{n+1}$.

The right-hand sides of Eqs.\ (\ref{eq:momentum}) and (\ref{eq:totenergy}) contain terms describing the effect of radiation pressure on the momentum and total energy densities, respectively, but are not included in the implicit radiation solve.  They are instead treated as explicit source terms and, using Eq.\ (\ref{eq:fick}), take the following forms:
\begin{align}
  \label{eq:mom_radpres}
  \frac{\rho\kappa_\mathrm{F}}{c}\vec{F}_\mathrm{R} =&~ -\lambda\nabla\er,\\
  \label{eq:etot_radpres}
  \frac{\rho\kappa_\mathrm{F}}{c}\vec{v}\cdot\vec{F}_\mathrm{R} =&~ -\lambda\vec{v}\cdot\nabla\er.
\end{align}
In practice, since $\er$ is a zone-centred quantity, Eq.\ (\ref{eq:mom_radpres}) is naturally face-centred and can therefore be applied to the momentum as is.  Eq.\ (\ref{eq:etot_radpres}) is also naturally face-centred, although it is applied to the zone-centred total energy.  As such, we adopt the following form (in Cartesian coordinates; \cf\ \citealt{zhangetal11}):
\begin{equation}
  \label{eq:et_radsrc}
  \begin{split}
  \lambda\vec{v}\cdot\nabla\er = \lambda_{i,j,k} \Bigg(&\, \frac{v_{1,i+1} + v_{1,i}}{2}\frac{\tilde{E}_{\mathrm{R}, i+1} - \tilde{E}_{\mathrm{R}, i}}{\Delta x_1}\\
  + & \frac{v_{2, j+1} - v_{2, j}}{2}\frac{\tilde{E}_{\mathrm{R}, j+1} - \tilde{E}_{\mathrm{R}, j}}{\Delta x_2}\\
  + & \frac{v_{3, k+1} + v_{3, k}}{2}\frac{\tilde{E}_{\mathrm{R}, k+1} - \tilde{E}_{\mathrm{R}, k}}{\Delta x_3} \Bigg),
  \end{split}
\end{equation}
where $\tilde{E}_\mathrm{R}^i$ denotes an upwinded interpolation of the radiation energy density in direction $i$ (see, \eg, \citealt{clarke96}).

The effects of radiation must also be included in the calculation of the time step.  Following \citet{krumholzetal07},
\begin{equation}
  \label{eq:dtraddiff}
  \Delta t_\mathrm{RD}^{-1} = \frac{1}{\Delta x_{\min}} \sqrt{\frac{4\er}{9\rho}\left(1 - e^{-\kappar\rho\Delta x_{\min}}\right)}
\end{equation}
is added to the existing time step calculation in quadrature to account for the (diffusion) radiation pressure.  Finally, the time step can be postdictively limited based on the maximum desired fractional change of $\er$ within a single step.  This is controlled via the user-defined tolerance \texttt{dttoler} (Table \ref{tab:parameterdescr}).
\subsection{Irradiation}
\label{sub:irrad}
The implementation of stellar irradiation in \azeus follows \citet{kuiperetal10,bitschetal13}, wherein the heating due to irradiation is added to the gas energy equation (Eq.\ \ref{eq:e_impsolv}) as a source term:
\begin{equation}
  \label{eq:nrg+irr}
  \frac{\del e}{\del t} = c\rho\left(\kappae \er - \kappap \arad T^4\right) + Q_\mathrm{irr}^{+}.
\end{equation}
With respect to solving the radiation system of equations, this manifests as an additional term in Eq.\ \ref{eq:t_np1}:
\begin{equation}
  \label{eq:t_np1_irr}
  T^{n+1} = \frac{T^n + 3 a_\mathrm{R} c \frac{\Delta t}{C_V} \kappa_\mathrm{P}^n (T^n)^4 + \frac{\Delta t}{C_V} c \kappa_\mathrm{E}^n E_\mathrm{R}^{n+1} + \frac{\Delta t}{\rho C_V} Q_\mathrm{irr}^{+} }{1 + 4 a_\mathrm{R} c \frac{\Delta t}{C_V} \kappa_\mathrm{P}^n (T^n)^3}.
\end{equation}

To include frequency-dependent transport of radiation in the irradiation algorithm, one simply sums over all frequency bins:
\begin{equation}
  \label{eq:sumqirrnu}
  Q_\mathrm{irr}^{+} = \int_0^{\infty}\!\! Q_{\mathrm{irr}, \nu}^{+}\, d\nu \approx \sum_{n' = 1}^{n} Q_{\mathrm{irr}, \nu}^{+} \Delta\nu(n'),
\end{equation}
where $n$ is the number of frequency bins, and $\Delta\nu(n')$ is the width of an individual frequency bin.  In the context of accretion disks and spherical coordinates, if the source is assumed to be at $r = 0$, the heating rate is thus given by:
\begin{equation}
  \label{eq:irrcases}
  Q_{\mathrm{irr}, \nu}^{+} =
  \begin{cases}
    \dfrac{\rho \kappa_{*,\nu}}{\Delta V_r} \int_0^{\Delta r}\! r^2 {F}_{\mathrm{irr}, \nu}\, dr & \text{if } \Delta\tau_\nu \leq \tau_\mathrm{thresh}; \\[1.5ex]
    \nabla\cdot (F_{\mathrm{irr}, \nu}\hat{\vec{r}}) & \text{otherwise},
  \end{cases}
\end{equation}
where $\kappa_{*, \nu}$ is the frequency-dependent opacity to stellar radiation, $\Delta\tau_\nu$ is the change in optical depth across $\Delta r$, $\Delta V_r = \frac{1}{3}(r^3_{i+1/2} - r^3_{i-1/2})$ is the radial ``volume element''\footnote{Whereas the total volume of a zone in spherical coordinates is given by $\Delta V = \Delta V_r \Delta V_\theta \Delta V_\phi = \frac{1}{3}\Delta(r^3)\Delta(-\cos\theta)\Delta\phi$.}, $F_{\mathrm{irr}, \nu}$ is the attenuated stellar flux:
\begin{equation}
  \label{eq:fluxirr}
  F_{\mathrm{irr}, \nu}(r) = F_{\mathrm{irr}, \nu}(R_*) \frac{R_*^2}{r^2} e^{-\tau_\nu(r)},
\end{equation}
$R_*$ is the stellar radius, $\tau_\nu(r) = \int_{R_*}^{\,r} \kappa_{*, \nu}\, \rho\, dr$ is the radial optical depth, and $F_{\mathrm{irr}, \nu}(R_*)$ is the flux at the stellar surface.  Although the two cases for the stellar heating rate in Eq.\ (\ref{eq:irrcases}) are mathematically equivalent, when calculating differences on a discrete grid, the differential form is subject to numerical round-off errors at very low optical depths \citep{brulsetal99}, and thus we use the integral form when $\Delta\tau_\nu \leq \tau_\mathrm{thresh}$, where typically $\tau_\mathrm{thresh} = 10^{-4}$.

Treating the stellar flux in this manner removes any explicit time-dependence, and thus it is valid only if the light travel time is shorter than the timescale for changes in the optical depth $\tau_\nu(r)$.  Conversely, this also means the calculation of the stellar flux at any location depends only on the instantaneous optical depth along a ray (and thus density and opacity), and can be decoupled from the hydrodynamics and diffuse radiation inside a single time step.  As such, the irradiation algorithm is invoked before solving the radiation matrix, and its results taken as constant during one combined hydrodynamic-radiation time step.

To obtain a grey version of the irradiation algorithm, the frequency-dependence in all quantities is dropped, and $\kappa_{*, \nu}$ is replaced by the Planck mean opacity at the stellar effective temperature, $\kappap(T_*)$.

The radiative force density due to stellar irradiation is also considered, and included as a source term in the momentum equation (Eq.\ \ref{eq:momentum}) akin to the radiation pressure term from FLD.  As we consider only single-group (grey) hydrodynamics, the irradiative flux is integrated over frequency before the force density is calculated:
\begin{equation}
  \label{eq:irrforce}
  a_\mathrm{irr} =
  \begin{cases}
    \dfrac{\rho \kappar(T_*)}{c\Delta V_r} \int_0^{\Delta r}\! r^2 {F}_\mathrm{irr}\, dr & \text{if } \Delta\tau \leq \tau_\mathrm{thresh}; \\[1.5ex]
    \dfrac{1}{c} \nabla\cdot ({F}_\mathrm{irr}\hat{\vec{r}}) & \text{otherwise}.
  \end{cases}
\end{equation}

The above discussion on irradiation assumes spherical coordinates and a disk-like setting.  However, the algorithm can be generalised to any curvilinear coordinate system and problem where the source irradiation propagates along coordinate axes.  Astrophysical examples of this could include the collected radiation from massive stars at a large distance from a region of interest, e.g., a protoplanetary disk or a photodissociation region.  Practically, the generalisation amounts to accounting for the appropriate geometric factors in Eqs.\ (\ref{eq:irrcases}) - (\ref{eq:irrforce}), and an example for Cartesian coordinates is presented in \S \ref{ssub:shadow}.
\subsection{Solving the radiation matrix}
\label{sub:solvematrix}
We solve Eq.\ (\ref{eq:er_differenced}) for $\er^{n+1}$ via a globally convergent Newton-Raphson method \citep{nr2nd}.  To begin, we re-write Eq.\ (\ref{eq:er_differenced}) in the following form:
\begin{equation}
  \label{eq:findroot}
  \begin{split}
  g_{i,j,k} = E^{n+1}_{\mathrm{R}, i,j,k} & - E^{n}_{\mathrm{R}, i,j,k}\\
  & + \Delta t c\rho^n \left[ \kappa^{n+1}_{\mathrm{E}, i,j,k} E^{n+1}_{\mathrm{R}, i,j,k} - \kappa^{n+1}_{\mathrm{P}, i,j,k} \arad (T^{n+1}_{i,j,k})^4 \right]\\
  & - \Delta t \nabla\cdot\left(D^n \nabla\er^{n+1}\right) - \Delta t (\tens{f}^{\,n}E^{n+1}_{\mathrm{R}, i,j,k})\!:\!\nabla\vec{v}^n,
  \end{split}
\end{equation}
where $T^{n+1}_{i,j,k}$ is given by Eqs.\ (\ref{eq:expandt_np1}) and (\ref{eq:t_np1}) (or [\ref{eq:t_np1_irr}] if irradiation is included), $\tens{f}^{\,n}$ is given by Eq.\ (\ref{eq:tenseddingfact}), the radiative diffusion term is described by Eq.\ (\ref{eq:radflux_differenced}), and the mean opacities are assumed to be functions of $T^{n+1}$.  This equation forms a sparse linear system of $N$ equations, where $N$ is the total number of zones on the grid, and is of the form:
\begin{align}
  \label{eq:linearsystem}
  & \mathcal{J}(\er)\, \delta\er = -g(\er),\\
  \intertext{where}
  \label{eq:jacobian}
  & \mathcal{J}(\er) = \sum_{n = 1}^N \frac{\del g_{i,j,k}}{\del E_{\mathrm{R},n}}
\end{align}
is the Jacobian, and $\delta\er$ is the change in the radiation energy density.  The desired solution of this system is $g_{i,j,k} = 0$.

Before attempting to solve Eq.\ (\ref{eq:linearsystem}), we first multiply each row of the matrix by the volume of its corresponding zone, $\Delta V = \Delta V_1 \Delta V_2 \Delta V_3$.  This symmetrises the matrix \citep{hayesetal06}, and thus we only need to calculate and store the diagonal and super-diagonal (or sub-diagonal) elements of the Jacobian.

Although \azeus includes tri-diagonal and successive over-relaxation (SOR) methods to solve Eq.\ (\ref{eq:linearsystem}), we instead primarily rely on the \petsc library \citep{petsc-web-page} due to its flexibility and performance.  Through experimentation, we find that the most generally reliable and efficient solver for our purposes is the iterative bi-conjugate gradient stabilised (BiCGSTAB) method plus a Jacobi (diagonal) preconditioner\footnote{For a list of the matrix solvers and preconditioners available in \petsc, see \url{http://www.mcs.anl.gov/petsc/documentation/linearsolvertable.html}.}, even though our radiation matrix is symmetric and permits the use of simpler solvers (such as the conjugate gradient method).

Convergence in the Newton-Raphson method is signalled when either $(\max |\delta\er^{(q)} / \er^{n+1,(q)}|)$ or $(\max |g_{i,j,k}^{(q)}|)$ fall below user-defined tolerances \texttt{nrtoler} and \texttt{nrtolrhs}, respectively, and where $(q)$ denotes the number of Newton-Raphson steps taken.  Furthermore, when using an iterative matrix solver, an additional user-defined tolerance is applied (\texttt{slstol}) to signal convergence with respect to the $L_2$-norm of the preconditioned residual ($r^{(p)} = -g - \mathcal{J} \delta \er^{(p)}$, where $(p)$ is the number of matrix solver iterations used).

At the end of the implicit solve, we calculate the change in the internal energy density ($\delta e^{n+1}$) via the ideal gas law and, if solving the total energy equation, subsequently update the total energy density using $\delta \etot^{n+1} = \delta e^{n+1} + \delta \er^{n+1}$.
\section{Considerations for AMR}
\label{sec:amr+rhd}
\subsection{Flux-limited diffusion}
\label{sub:amr+fld}
We have chosen to implement AMR + FLD using the ``deferred synchronisation'' algorithm of \citet{zhangetal11}.  This approach derives from the algorithm of \citet{howellgreenough03}, but does not require a simultaneous matrix solution over multiple levels of the AMR hierarchy for flux synchronisation, thus reducing the complexity of the algorithm and improving performance.

In the AMR algorithm, at the end of $\xi$ time steps at level $l+1$, where $\xi$ is the refinement ratio (and typically equal to 2), zones at level $l$ which lie under zones at level $l + 1$ are conservatively overwritten using the higher resolution data.  This introduces differences along the edges of level $l+1$ grids because zones at level $l$ were evolved using fluxes calculated on level $l$, but these are generally not the same as fluxes calculated at the same location on a grid at level $l+1$.  To restore consistency and, more importantly, conservation, fluxes calculated at level $l$ that are co-spatial with the edges of grids at level $l+1$ are replaced by the higher resolution fluxes.  Operationally, these replacements are applied directly to the (M)HD variables at level $l$ as so-called ``flux corrections'' before another time step is taken.  This process of replacement and correction is often called ``restriction''.

In the deferred synchronisation algorithm, radiative fluxes ($\fr$; Eq.\ \ref{eq:fick}) calculated during an implicit radiation solve are saved along the edges of each grid at level $l+1$, analogous to the fluxes for the (M)HD quantities.  While corrections to the (M)HD variables at level $l$ are applied explicitly as described above, the radiative flux corrections are instead applied during the next implicit radiation solve as a source term on the right-hand side of the radiation matrix.

For example, in Cartesian coordinates and the 1-direction, the radiative flux correction for a coarse zone $(I,J,K)$ at level $l$ can be written in terms of a sum in time and space over the underlying fine radiative fluxes\footnote{Fine zone indices $(i,j,k)$ correspond to the (left, bottom, back) of coarse zone $(I,J,K)$, a given fine zone centre is denoted using $(i+\alpha,j+\beta,k+\eta)$, where $\alpha, \beta, \eta = 0,\ldots, \xi-1$, and a 1-component of flux is then denoted with $(i,j+\beta,k+\eta)$ and $\beta, \eta = 0,\ldots, \xi-1$.}:
\begin{equation}
  \label{eq:radflxcorr}
  \delta {F}_{\mathrm{R},1}^{N+1}(I,J,K) = {F'}_{\mathrm{R},1}^{N+1}(I,J,K) - \sum_{\beta, \eta, \tau=0}^{\xi - 1} {f'}_{\mathrm{R},1}^{n+\tau+1}(i,j+\beta,k+\eta),
\end{equation}
where ${F'}_{\mathrm{R},1}^{N+1}$ is the coarse radiative 1-flux (with units $F_\mathrm{R}\Delta A \Delta t$, where $\Delta t$ and $\Delta A = \Delta x_2 \Delta x_3$ are the coarse time step and area element, respectively)\footnote{Although the inclusion of $\Delta A$ and $\Delta t$ in the definition means this is not strictly a flux but rather an energy, we find it advantageous for the AMR bookkeeping to define all fluxes in this manner.}, and ${f'}_{\mathrm{R},1}^{n+\tau+1}$ are the corresponding fine radiative 1-fluxes at level $l+1$ (with units $f_\mathrm{R}\delta A \delta t$, where $\delta t$ and $\delta A = \delta x_2 \delta x_3$ are the fine time step and area element, respectively).

Once the radiative flux corrections are known, Eq.\ (\ref{eq:radflxcorr}) and the equivalent expressions in the 2- and 3-directions are added to the right-hand side of the matrix equation (Eq.\ \ref{eq:findroot}) to give:
\begin{align}
  \label{eq:findroot_amr}
  \begin{split}
  g_{i,j,k} = \er^{n+1} - \er^{n} & + \Delta t c\rho^n \left[ \kappae^{n+1} \er^{n+1} - \kappap^{n+1} \arad (T^{n+1})^4 \right]\\
  & - \Delta t \nabla\cdot\left(D^n \nabla\er^{n+1}\right) - \Delta t (\tens{f}^{\,n}\er^{n+1})\!:\!\nabla\vec{v}^n\\
  & + \frac{1}{\Delta V(I,J,K)}\delta{F}_\mathrm{R}^{N+1}(I,J,K);
  \end{split}\\
  \delta{F}_\mathrm{R}^{N+1} = &~ \delta{F}_{\mathrm{R},1}^{N+1} + \delta{F}_{\mathrm{R},2}^{N+1} + \delta{F}_{\mathrm{R},3}^{N+1}.\nonumber
\end{align}
The solution of the radiation matrix then proceeds as in \S \ref{sub:solvematrix}.

In the absence of radiation, at the end of a time step at level $l$, and after the flux corrections have been applied to the hydrodynamic variables, all grids at levels $l \rightarrow l_{\max}$ are co-temporal and consistent.  As discussed in detail in \citet{zhangetal11}, when applying the radiative flux corrections using deferred synchronisation, consistency no longer holds at the end of a time step.  Although this does not affect results in practice, it does create three complications.  First, any time grids at level $l' > l$ are adjusted, created, or destroyed (the ``regridding'' step), in order to maintain consistency, any existing radiative flux corrections must carry over to the modified grid structure for application during the next implicit radiation solve at level $l$.  Second, when adjacent refined ($l' > l$) grids are present, radiative flux corrections should only be applied to zones on grids at level $l$ where there are no overlying refined zones.  To do so is to ``doubly-correct'' the coarse zone (by overwriting and by flux correction).  For the (M)HD variables, this never poses a problem because the overwriting takes place after the flux corrections, in contrast to the deferred synchronisation algorithm.  Finally, checkpoints are created at the end of a level $l = 1$ time step when all levels are co-temporal, and the deferred radiative flux corrections must be included in the checkpoints to ensure consistency upon a restart.
\subsection{Irradiation}
\label{sub:amr+irr}
Coupling the ray-tracing algorithm described in \S \ref{sub:irrad} to the AMR is relatively simple and straightforward.  First, we interpolate the irradiative fluxes $F_\mathrm{irr}(I,J,K)$ from level $l$ to any corresponding overlying boundary zones at level $l+1$ under the following requirement:
\begin{equation}
  \label{eq:fluxinterp}
  \begin{split}
  F_\mathrm{irr}(I,J,K)\Delta A(I,J,K) = \sum_{\beta, \eta=0}^{\xi - 1} f_\mathrm{irr}&(i,j+\beta,k+\eta)\\[-2.0ex]
  & \times \delta A(i,j+\beta,k+\eta),
  \end{split}
\end{equation}
where $F_\mathrm{irr}$ and $f_\mathrm{irr}$ have the canonical definitions of flux (in contrast to ${F'}_{\mathrm{R},1}$ and ${f'}_{\mathrm{R},1}$ in Eq.\ \ref{eq:radflxcorr}), and $\Delta A$, $\delta A$ are defined as before.  If there is instead an adjacent or overlapping grid at the same level, data is then taken directly from this grid.  This is the same procedure as applied when interpolating other face-centred quantities over a surface area in \azeus (\eg, the magnetic field; see \citealt{rcm12} for details), and is generally known as ``prolongation''.

The interpolated values are then used as input (along with the local optical depth) to the ray-tracing algorithm to determine irradiation fluxes in the active zones of each grid at level $l+1$\footnote{By active zones, we mean zones to which the equations of RHD are self-consistently applied; \ie, not boundary zones.}, followed by the irradiative heating.  At the end of a time step at level $l$, the RHD variables are overwritten as described in \S \ref{sub:amr+fld}, excepting the irradiative fluxes.  Since the irradiative flux at any point is globally determined, and is recalculated in the active zones at the beginning of every time step, it would be pointless to perform a restriction step.  Rather, it is the consistency of the local optical depth $\Delta\tau$, \ie, the density and opacity, that is important.

We emphasise that the constraint on the irradiation algorithm discussed earlier remains: the time scale for changes in the optical depth must be longer than the light travel time scale.  Further, when coupling the ray-tracing algorithm to AMR, this constraint extends to \emph{all} levels.  In other words, because a particular value of the irradiative flux can depend on $\Delta\tau$ from a different grid, and which may have a different resolution, in order to maintain accuracy, changes in $\Delta\tau$ must occur on time scales comparable to or less than the smallest time step in the simulation.

\section{Numerical tests}
\label{sec:tests}
Below we present several test problems intended to validate different aspects of our radiation hydrodynamics algorithms.  Unless otherwise noted, the BiCGSTAB linear solver and Jacobi/diagonal preconditioner from \petsc is used with a relative tolerance of \texttt{slstol} $ = 10^{-6}$, and the total energy equation is employed.  A description of the most important user parameters, as well as the values used herein, can be found in Tables \ref{tab:parameterdescr} and \ref{tab:testparameters}. For tests of the MHD and/or AMR algorithms in \azeus, we direct the reader to \citet{clarke10} and \citet{rcm12}.

\begin{table*}[htb]
  \centering
  \caption{Description of important user-set parameters in \azeus.}
  \begin{tabular}{lll}
  \hline\hline
  Parameter name & Description & Range \\
  \hline
  \texttt{courno} & the Courant number & 0 -- 1\\
  \texttt{qcon} & quadratic artificial viscosity parameter\tablefootmark{a} & $\geq$ 0 \\
  \texttt{qlin} & linear artificial viscosity parameter\tablefootmark{a} & $\geq$ 0 \\
  \texttt{ibuff} & number of buffer zones placed around a new grid & $\geq$ 0 \\
  \texttt{geffcy} & minimum allowed fractional grid efficiency for creation of a new grid & 0 -- 1 \\
  \texttt{kcheck} & number of cycles at a level between refinement checks & $\geq$ 0 \\
  \texttt{nrtoler} & tolerance for max.\ relative change in $\er$ during one call to the radiation solver & $>$ 0\\
  \texttt{nrtolrhs} & tolerance for max.\ magnitude of the matrix RHS during one call to the radiation solver & $\geq$ 0\\
  \texttt{slstol} & relative tolerance of the iterative matrix solver & $>$ 0 \\
  \texttt{dttoler} & maximum \emph{desired} relative change in $\er$ during one call to the radiation solver & $>$ 0\\
  \hline
  \end{tabular}
  \tablefoot{\tablefoottext{a}{We refer the reader to \citet{clarke10} for a detailed description of the artificial viscosity parameters.}
  }
  \label{tab:parameterdescr}
\end{table*}

\begin{table*}
  \caption{Values employed for important user-set parameters in \S \ref{sec:tests}.}
  \centering
  \begin{tabular}{lllllllllll}
  \hline\hline
  Test problem & \texttt{courno} & \texttt{qcon} & \texttt{qlin} & \texttt{ibuff} & \texttt{kcheck} & \texttt{geffcy} & \texttt{nrtoler} & \texttt{nrtolrhs} & \texttt{dttoler}\\
  \hline
  \rule{0pt}{2ex}
  \hyperref[ssub:lindiff]{Linear diffusion} & 1 & ... & ... & 1 & 5 & 1.0 & $10^{-8}$ & $10^{-13}$ & 0.2 \\
  \hyperref[ssub:coupling]{Radiation-matter coupling} & 1 & ... & ... & ... & ... & ... & $10^{-8}$ & $10^{-6}$ & 1 \\
  \hyperref[ssub:marshak]{Marshak waves} & 1 & ... & ... & ... & ... & ... & $10^{-8}$ & $10^{-10}$ & 1 \\
  \hyperref[ssub:radshock]{Radiative shock waves} & 0.5 & 1.0 & 0.1 & 3 & 30 & 1.0 & $10^{-8}$ & $10^{-6}$ & 0.3 \\
  \hyperref[ssub:diskrelax]{PPD relaxation benchmarks}\!\! & 0.5 & 0 & 0 & ... & ... & ... & $10^{-4}$ & $10^{-4}$ & 0.5 \\
  \hyperref[ssub:staticdiskrt]{static disk RT benchmarks} & 1 & ... & ... & ... & ... & ... & $10^{-4}$ & $10^{-11}$ & 0.3 \\
  \hyperref[ssub:shadow]{Shadow test (static)} & 0.5 & 2.0 & 0 & 4 & 30 & 0.8 & $10^{-4}$ & $10^{-4}$ & 1.0 \\
  \hyperref[fig:shadowevap]{Shadow test (dynamic)} & 0.5 & 2.0 & 0 & 5 & 30 & 0.5 & $10^{-4}$ & $10^{-4}$ & 1.0 \\
  \hline
  \end{tabular}
  \label{tab:testparameters}
\end{table*}
\subsection{1-D: Linear diffusion}
\label{ssub:lindiff}
The first test we present follows \citet{commerconetal11,commerconetal14,kolbetal13}, and models a radiative energy pulse as it diffuses outward; all other physical modules are deactivated.  Thus, the equation being solved is:
\begin{equation}
  \label{eq:purediff}
  \frac{\del E_\mathrm{R}}{\del t} - \nabla\cdot\left( \frac{c}{3\rho\kappa_\mathrm{R}}\nabla E_\mathrm{R}\right) = 0,
\end{equation}
where $\lambda = 1/3$ (the Eddington approximation; no flux limiting), corresponding to an optically thick regime.  

We set $\rho \kappa_\mathrm{R} = 1$, which then gives the following analytical solution to Eq.\ (\ref{eq:purediff}) in 1-D:
\begin{equation}
  \label{eq:lindiff_soln}
  E_\mathrm{R}(x,t) = \tilde{E}_o \frac{1}{2\! \sqrt{D \pi t}}\,e^{-x^2 / 4 D t},
\end{equation}
where $D = c / ( 3 \rho \kappa_\mathrm{R} )$ is the diffusion coefficient, and $\tilde{E}_o = E_\mathrm{R}(x = 0, t = 0) \Delta x = 10^5$ is the initial value of the energy pulse.  Under these assumptions, this test is equivalent to linear thermal conduction (\eg, \citealt[\S 103]{mm84}).  Similar to linear conduction, a diffusion coefficient which is independent of temperature and greater than zero everywhere results in an infinite signal propagation speed, and the energy pulse will diffuse at speeds greater than the speed of light.

A Cartesian domain of size $x \in [-0.5, 0.5]$ is employed, and the pulse is initialised on the coarsest level ($l = 1$) as a delta function at $x = 0$.  In \azeus, because the refinement ratio must be a power of two, and because interpolations do not introduce new extrema, the initial delta function thus becomes a finite width pulse at higher resolutions, but always satisfies $\tilde{E}_o = E_\mathrm{R} \Delta x_{l = 1}$.  The coarsest level has 33 zones, a refinement ratio of 2 and 3 levels of refinement are used, giving an effective resolution of 264 zones. As in \citet{commerconetal14}, zones are flagged for refinement when the gradient in the radiation energy density exceeds 25\%, and the time step on the coarsest level is maintained at $3.125\expon{-15}\,\mathrm{s}$.  For a summary of the parameters used in this test, as well as others, see Table \ref{tab:testparameters}.

\begin{figure}[htb]
  \resizebox{\hsize}{!}{\includegraphics{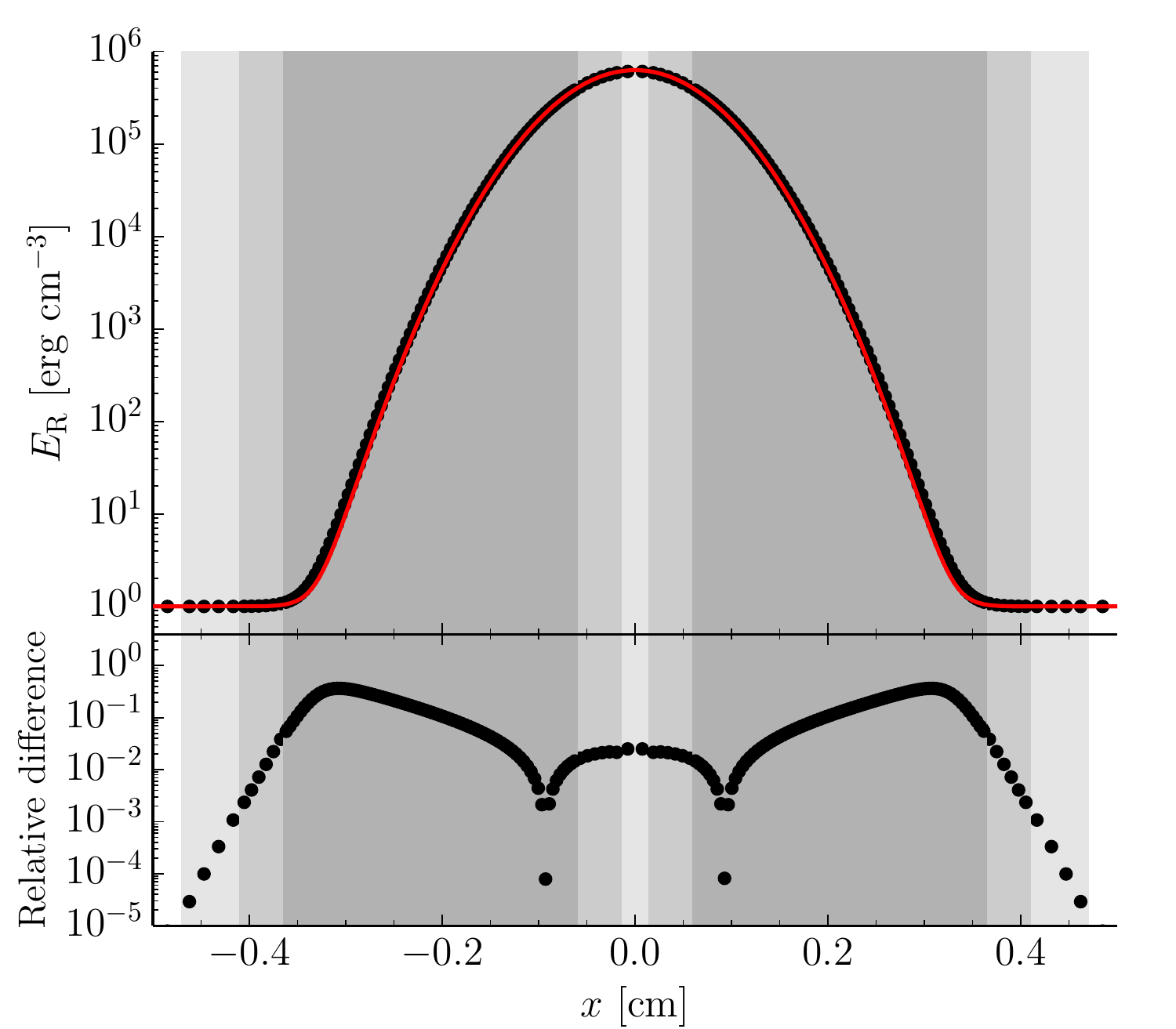}}
  \caption{Linear diffusion test.  \textit{Top:} the radiation energy density at $t = 2\expon{-13}\,\mathrm{s}$ (black), and corresponding analytical solution (red).  \textit{Bottom:} the relative difference between analytical and numerical solutions.  Successively darker shading indicates higher levels of refinement (from $l = 1 \rightarrow 4$).}
  \label{fig:lindiff}
\end{figure}

Figure \ref{fig:lindiff} shows the results at $t = 2\expon{-13}\,\mathrm{s}$, as well as the fractional relative difference from the analytical solution (Eq.\ \ref{eq:lindiff_soln}).  The numerical result matches well with the analytical solution, and agrees with the results of \citet{commerconetal14}.  The large differences toward the edges of the diffusion front are a result of the first-order backwards Euler implicit method; indeed, the relative differences in this case do not vary strongly with spatial resolution.  It should be noted that, while the relative difference always remains quantitatively similar to a uniform grid simulation of the same effective resolution, the shape of the relative difference profile depends noticeably on the specific values of the refinement criteria (\eg, \texttt{ibuff}, \texttt{geffcy}, \etc).
\subsection{1-D: Radiation-matter coupling test}
\label{ssub:coupling}
This standard benchmark was originally presented in \citet{turnerstone2001}, and tests the coupling between radiation and fluid energies.  A stationary, uniform combination of radiation and fluid is used, but it is initially out of equilibrium.  The radiation energy density is initialised to $10^{12}\, \mathrm{erg\, cm}^{-3}$, and is assumed dominant relative to the fluid energy, and therefore roughly constant in time.  Under these assumptions, Eqs.\ (\ref{eq:e_impsolv})--(\ref{eq:er_impsolv}) decouple, and the system simplifies to a single ordinary differential equation:

\begin{equation}
  \label{eq:gasnrg_ode}
  \frac{d e}{d t} = c\kappap\rho\left( \er - a_\mathrm{R}T^4\right)
\end{equation}

Given a constant density and Planck opacity, plus the ideal gas law, this equation can be numerically integrated to determine a reference solution.

Two versions of the test are presented using very different initial internal energy densities: $10^2\, \mathrm{erg\, cm}^{-3}$ and $10^{10}\, \mathrm{erg\, cm}^{-3}$, but the same radiation energy density ($10^{12}\, \mathrm{erg\, cm}^{-3}$).  The density and Planck mean opacity are set to $\rho = 10^{-7}\, \mathrm{g\, cm}^{-3}$ and $\kappap = 0.4\, \mathrm{cm^2\, g^{-1}}$, respectively, while the ratio of specific heats is set to $\gamma = 5/3$, and the mean molecular weight is set to $\mu = 0.6$.  To ensure good time resolution in the plotted results, we follow \citet{kolbetal13} and set our initial time step to $10^{-20}\,\mathrm{s}$, and then allow it to increase by 5\% each subsequent cycle until $t = 10^{-4}\,\mathrm{s}$ is reached.  As the conditions should remain uniform, this test should be independent of numerical resolution, but for completeness, we employ a 1-D grid with a resolution of 16 zones in Cartesian coordinates.  AMR is not used for this test, neither is a flux limiter ($\lambda = 1/3$), and all other physical modules are deactivated.

\begin{figure}[htb]
  \resizebox{\hsize}{!}{\includegraphics{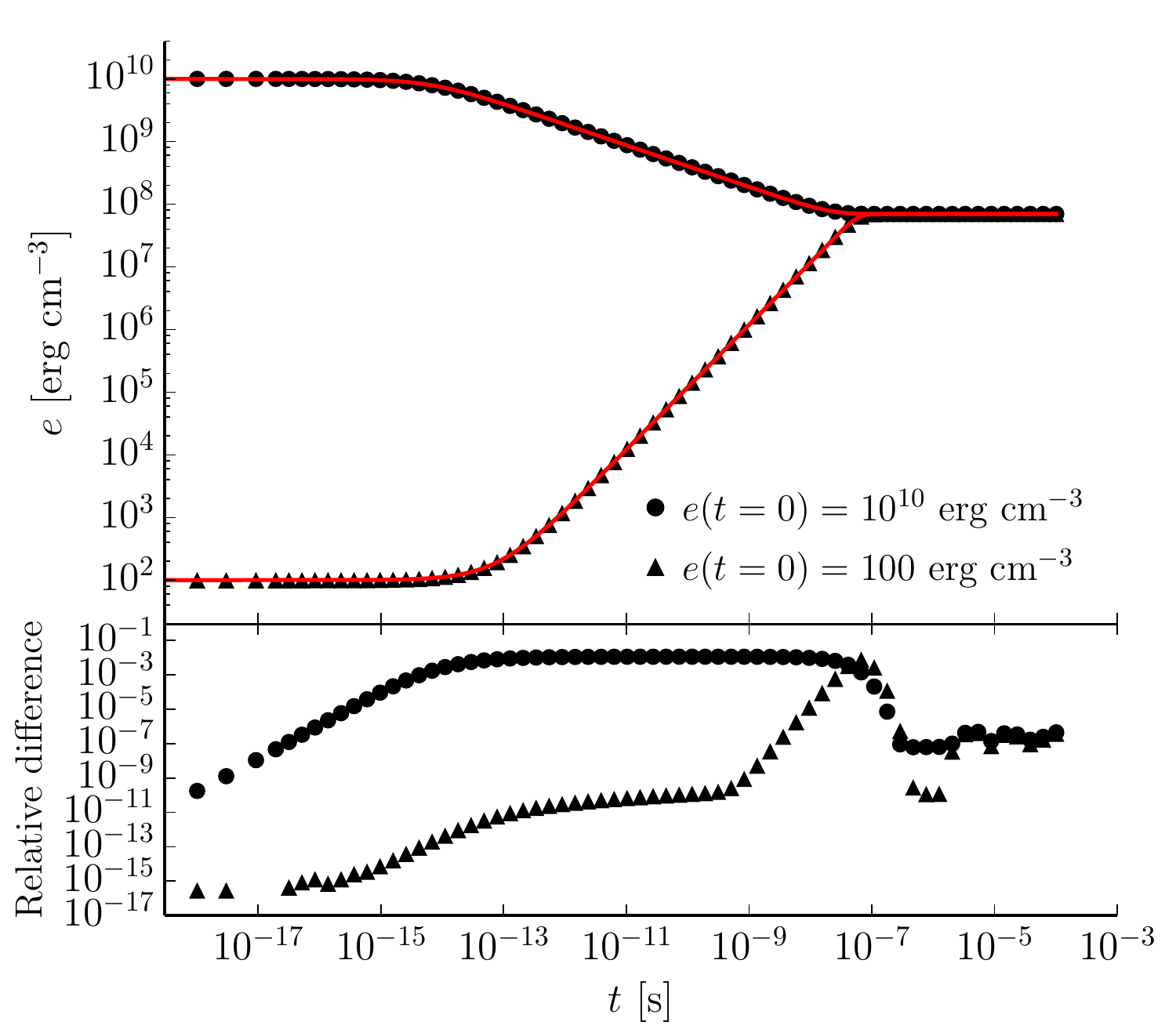}}
  \caption{Radiation-matter coupling test.  \textit{Top:} the internal energy density as a function of time given two different initial states: $10^2\, \mathrm{erg\, cm}^{-3}$ (triangles) and $10^{10}\, \mathrm{erg\, cm}^{-3}$ (circles).  The reference solution is also plotted (red). \textit{Bottom:} the relative difference between numerical and reference solutions.}
  \label{fig:coupling}
\end{figure}

Figure \ref{fig:coupling} plots the numerical results for the internal energy density as a function of time, the reference solution, as well as the relative difference between the two.  For a time step growth rate of 5\% per cycle, the relative difference between numerical and reference solutions has a maximum of $\sim\! 10^{-2}$, while the relative difference of the final solution fluctuates around $2\times 10^{-7}$.  We also find the maximum relative difference decreases roughly linearly with decreasing growth factor, as would be expected for the first-order backwards Euler method used here.  It is, however, worthwhile to note that, even with a growth rate of 50\%, we still obtain a final solution to within a relative difference of $5\times 10^{-7}$ .
\subsection{1-D: Marshak waves}
\label{ssub:marshak}
The non-equilibrium Marshak wave test is a time-dependent non-linear diffusion test for which there exists analytic solutions \citep{suolson96}.  For this test, consider an initially zero temperature, homogeneous, static, and semi-infinite medium which is illuminated on one side by some external radiation flux $F_\mathrm{inc}$.  The material is assumed to have a constant opacity $\kappa$, but a specific heat capacity at constant volume given by $C_V = \alpha T^3$, where $\alpha$ is a parameter.  With a specific heat capacity of this form, Eqs.\ (\ref{eq:e_impsolv})--(\ref{eq:er_impsolv}) become linear in $\er$ and $T^4$.

Following \citet{pomraning1979}, we adopt the following dimensionless variables:
\begin{align}
  \label{eq:marshak_x}
  x \equiv&~ \sqrt{3}\kappa z;\\
  \label{eq:marshak_t}
  \tau \equiv&~ \left(\frac{4\arad c \kappa}{\alpha}\right) t;\\
  \label{eq:marshak_u}
  u(x,\tau) \equiv&~ \frac{c}{4}\left(\frac{\er (z,t)}{F_\mathrm{inc}}\right);\\
  \label{eq:marshak_v}
  v(x,\tau) \equiv&~ \frac{c}{4}\left(\frac{\arad T^4}{F_\mathrm{inc}}\right).
\end{align}

The initial and boundary conditions, given in dimensionless variables, are:
\begin{align}
  \label{eq:marshak_bc1}
  &u(0,\tau) - \frac{2}{\sqrt{3}}\frac{\del u(0,\tau)}{\del x} = 1,\\
  \label{eq:marshak_bc2}
  &u(\infty,\tau) = u(x,0) = v(x,0) = 0.
\end{align}

The additional parameter, $\epsilon = 4\arad / \alpha$, is set to $0.1$ to permit direct comparison with the solutions of \citet{suolson96}.

A specific heat capacity $C_V \propto T^3$ presents a complication for the FLD algorithm presented here:  Eq.\ (\ref{eq:temp_impsolv}), and consequently Eqs.\ (\ref{eq:temp_differenced}) \& (\ref{eq:t_np1}), assume that $C_V$ is independent of temperature, and thus solving the radiation matrix as described in \S \ref{sub:fld} will give incorrect results for this test.

However, the gas energy equation is easily modified to correct for this.  From Eq.\ (\ref{eq:e_impsolv}) and the ideal gas law, we can write:
\begin{equation}
  \label{eq:temp_impsolv_marshak}
  \frac{\del (C_V T)}{\del t} = \alpha\frac{\del (T^4)}{\del t} = c\left(\kappa_\mathrm{E} \er - \kappa_\mathrm{P} \arad T^4\right),
\end{equation}
from whence an expression for the updated temperature $T^{n+1}$ can be derived, without the need to linearise any terms:
\begin{equation}
  \label{eq:t_np1_marshak}
  (T^{n+1})^4 = \frac{\alpha (T^n)^4 + c \Delta t^n \kappae^n E_\mathrm{R}^{n+1} }{\alpha + c \Delta t^n \kappap^n \arad}.
\end{equation}
Eq.\ (\ref{eq:t_np1_marshak}) replaces Eq.\ (\ref{eq:t_np1}) in the implicit radiation solver.

\begin{figure}[htb]
  \resizebox{\hsize}{!}{\includegraphics{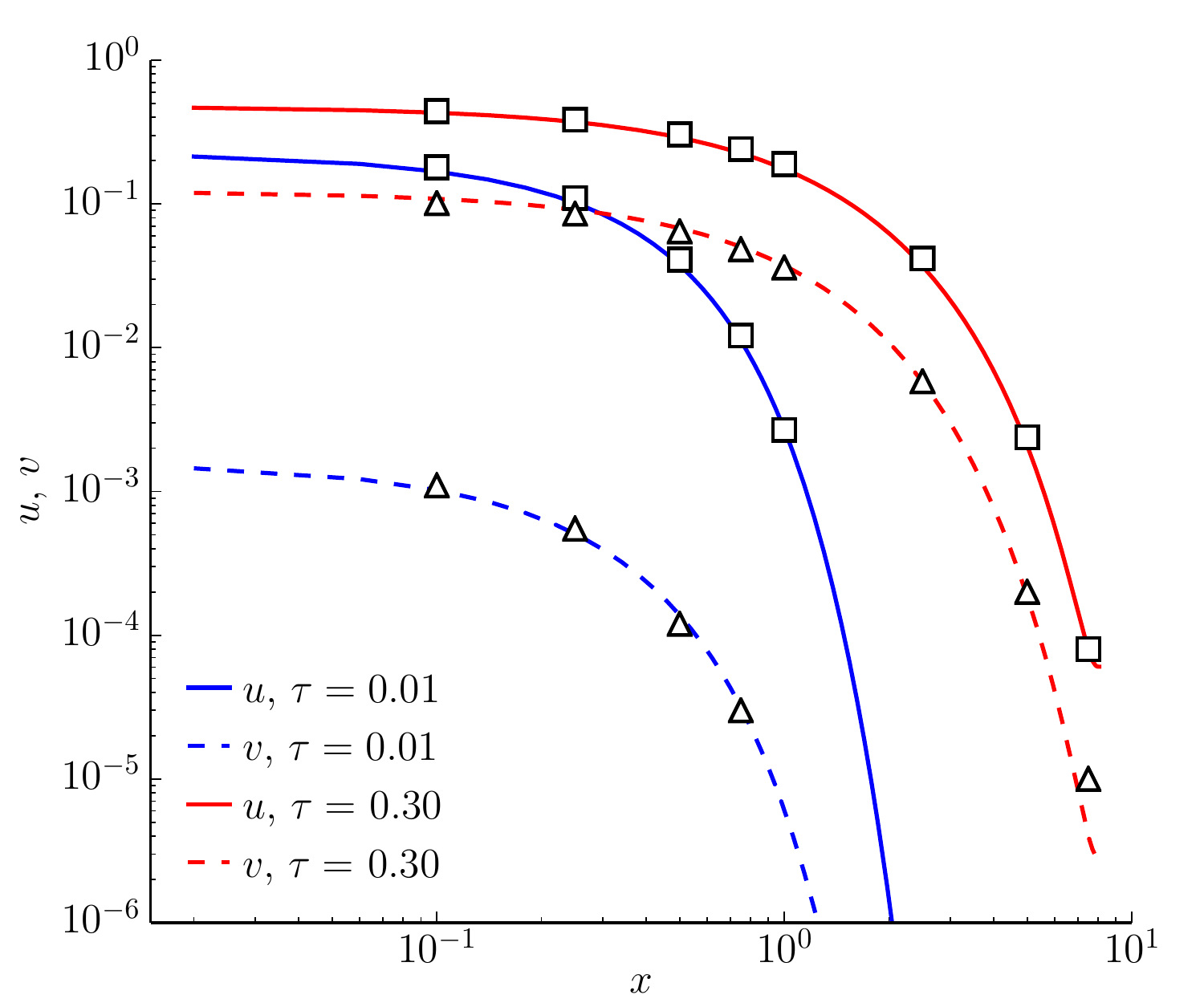}}
  \caption{Marshak wave test.  The dimensionless radiation energy density $u$ and gas energy density $v$ are plotted at times $\tau = 0.01$ and $\tau = 0.3$.  Numerical results are shown as curves, while reference data are shown as symbols (squares: $u$; triangles: $v$).}
  \label{fig:marshak}
\end{figure}

Following \citet{hayesetal06}, we model a 1-D Cartesian domain with 200 zones and length 8 cm.  The density is set to $\rho = 1\, \mathrm{g\, cm^{-3}}$, the opacity to $\kappa = 1 / \! \sqrt{3}$, and the dimensionless time step to $\Delta\tau = 3 \times 10^{-4}$ \citep{zhangetal11}.  As in the previous test, AMR is not used, $\lambda = 1/3$, and all other physical modules are deactivated.

Figure \ref{fig:marshak} plots the numerical results for the dimensionless radiation energy and internal energy densities as a function of $x$ at two different times: $\tau = 0.01$ and $\tau = 0.3$.  The corresponding benchmark data of \citet{suolson96} is also plotted.  The agreement between numerical and analytical data is very good.

\subsection{1-D: Radiative shock waves}
\label{ssub:radshock}
An important test for any RHD code are radiative shock tubes, where the hydrodynamic shock structure is substantially modified by the presence of radiation (\eg, \citealt{mm84}).  Determining analytical solutions for this problem is not possible in general, but \citet{lowrieedwards08} have calculated piece-wise semi-analytic solutions for the non-equilibrium diffusion case, and which are suitable for comparison with numerical results.  

\begin{figure*}[htb]
  \resizebox{0.99\hsize}{!}{\includegraphics{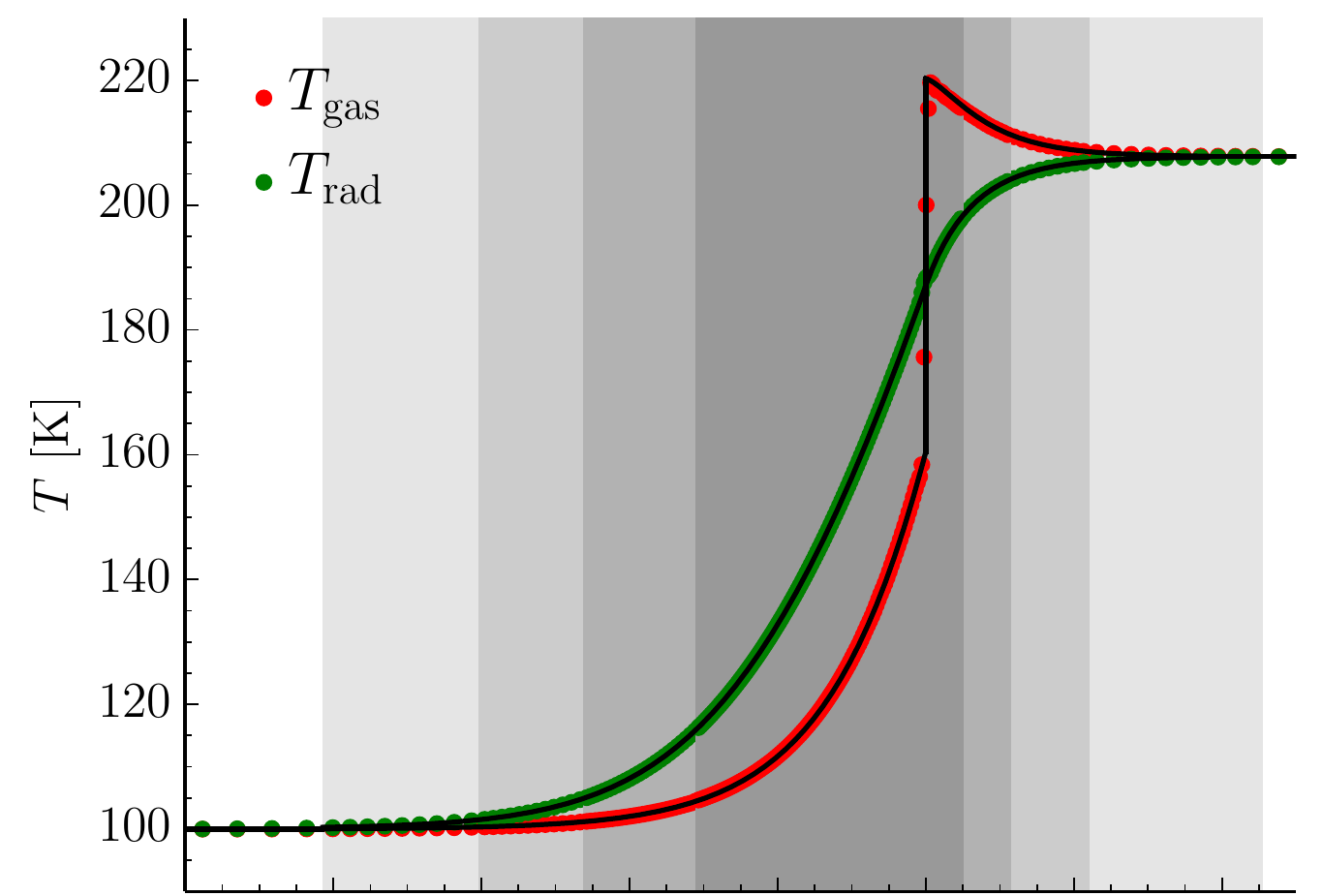}\qquad\includegraphics{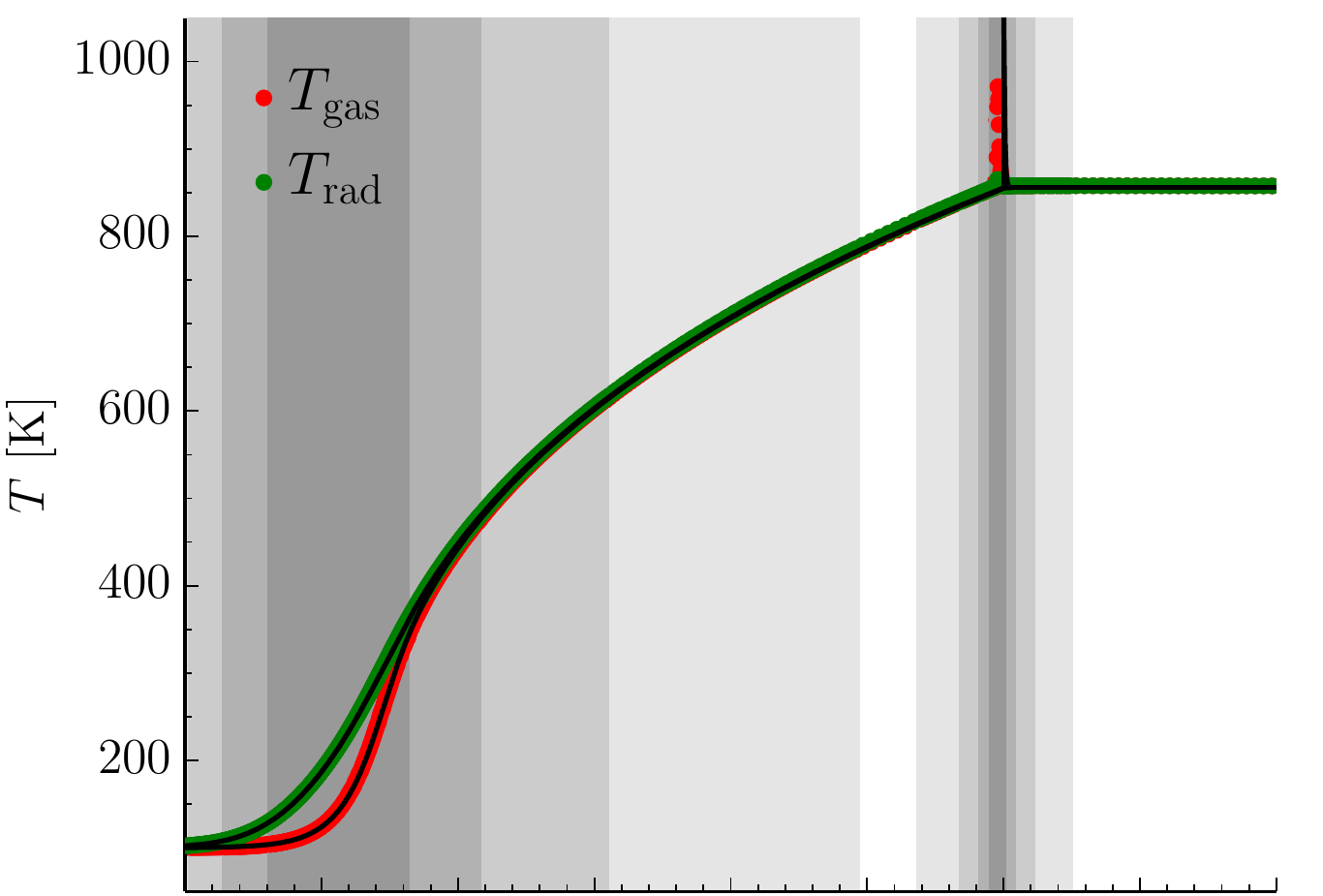}}
  \resizebox{0.99\hsize}{!}{\includegraphics{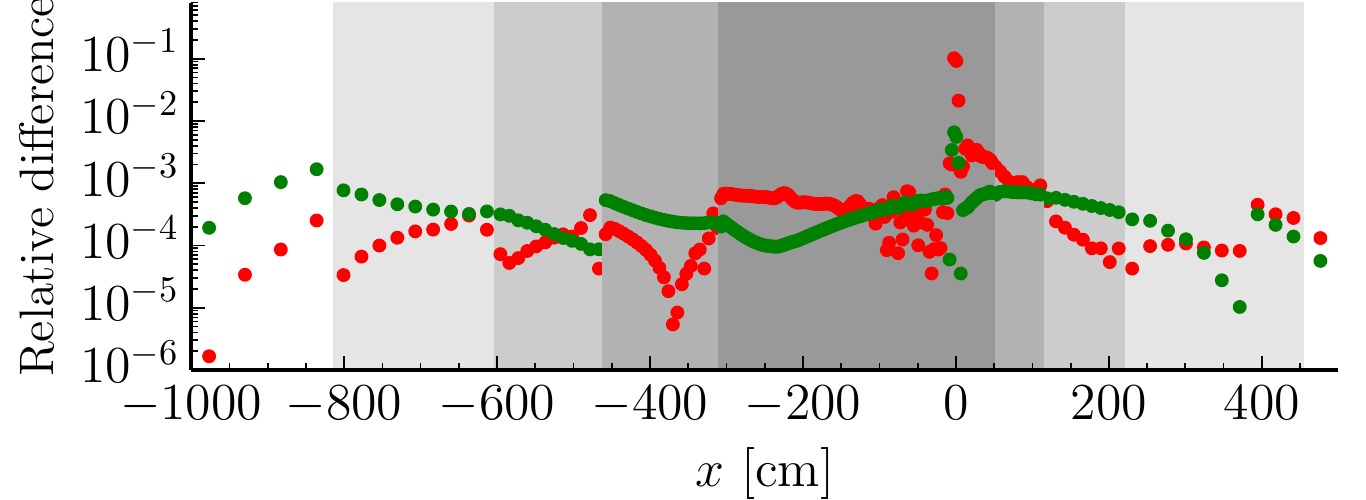}\qquad\includegraphics{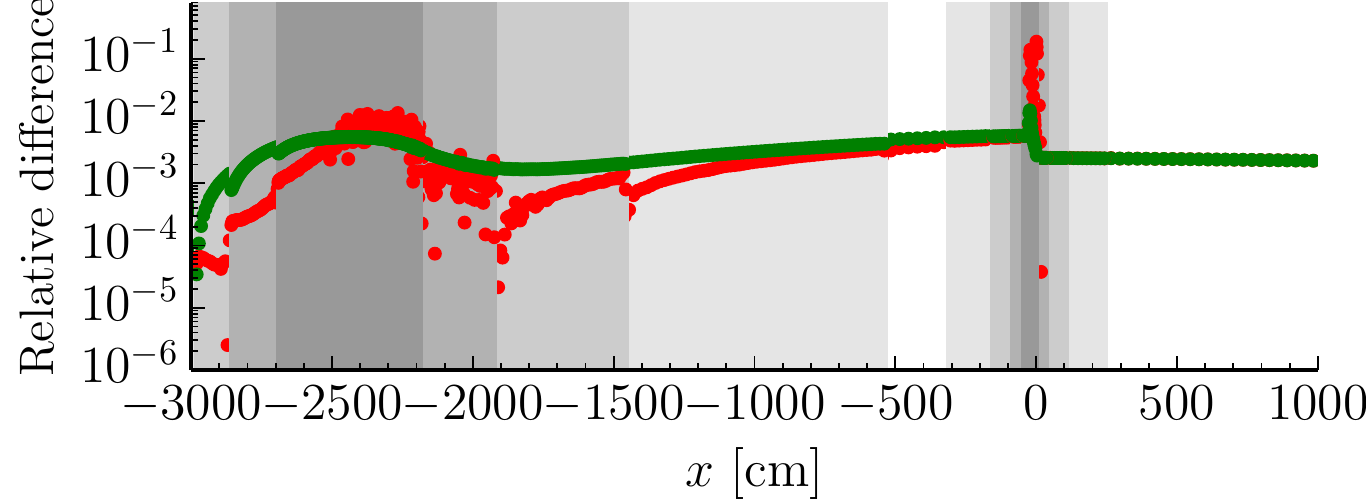}}
  \caption{Radiative shock wave tests. \textit{Top:} gas (red) and radiation (green) temperatures in $M = 2$ sub-critical (\textit{left}) and $M = 5$ super-critical (\textit{right}) radiative shocks at $t = 0.1\,\mathrm{s}$, plus corresponding semi-analytic solutions (black). \textit{Bottom:} the relative difference between numerical and semi-analytical solutions.  As before, successively darker shading indicate higher levels of refinement (from $l = 1 \rightarrow 5$).}
  \label{fig:radshk}
\end{figure*}

Radiative shock waves generally fall into two categories: sub- and super-critical.  In the first case, radiation from the post-shock material diffuses upstream, heating the gas, although the temperature of this ``precursor'' remains significantly below the post-shock temperature.  Note that gas and radiation are out of equilibrium in this region, and that its extent depends primarily on the opacity of the gas.   Meanwhile, precursor and downstream states are connected by an embedded hydrodynamic shock and a subsequent relaxation region that eventually asymptotes to downstream values.  

Following \citet{zhangetal11}, we initialise a stationary sub-critical $M = 2$ shock tube in 1-D, with domain $-1000 \leq x \leq 500\, \mathrm{cm}$, and two constant states separated by a discontinuity at $x = 0$.  The left (upstream) state has the following properties: $\rho_0 = 5.45887 \times 10^{-13}\, \mathrm{g\, cm^{-3}}$, $T_0 = 100\, \mathrm{K}$, and $v_0 = 2.35435 \times 10^5\, \mathrm{cm\, s^{-1}}$, while the right (downstream) state is given by: $\rho_1 = 1.24794 \times 10^{-12}\, \mathrm{g\, cm^{-3}}$, $T_1 = 207.757\, \mathrm{K}$, and $v_1 = 1.02987 \times 10^{5}\, \mathrm{cm\, s^{-1}}$.  Gas and radiation are initially assumed to be in equilibrium ($T = T_\mathrm{R}$; $T_\mathrm{R} = [\er / \arad]^{1/4}$ is the radiation temperature).  The Planck and Rosseland extinction coefficients are set to $\alpha_\mathrm{P} = \rho \kappap = 3.92664 \times 10^{-5}\, \mathrm{cm^{-1}}$, and $\alpha_\mathrm{R} = \rho \kappar = 0.848902\, \mathrm{cm^{-1}}$, respectively, and we use $\gamma = 5/3$, $\mu = 1$.  Fixed boundary conditions are applied on either side of the domain with values given by the initial left and right states.  Following \citet{lowrieedwards08}, we apply the Eddington approximation ($\lambda = 1/3$).  For the AMR, the coarsest level has a resolution of 32 zones, and 4 levels of refinement are employed, resulting in an effective resolution of 512 zones.  Zones are flagged based on gradients in the internal and total energy densities, at a threshold of 5\%.  As in \citet{zhangetal11,commerconetal14}, we find that the initial discontinuity must be shifted slightly to ensure the final steady-state position of the shock lies at $x = 0$.  For the sub-critical case, an offset of $-100$ cm was required, but we note that the specific value of this shift is sensitive to the values of the artificial viscosity parameters (\texttt{qcon}, \texttt{qlin}), the Courant number, and the numerical resolution.

The left side of Figure \ref{fig:radshk} plots the gas and radiation temperatures at $t = 0.1\, \mathrm{s}$ for the sub-critical $M = 2$ case, along with the relative differences from the semi-analytic solution of \citet{lowrieedwards08}.  With the exception of the region near the embedded shock, where the semi-analytical solution contains a true discontinuity, the agreement is excellent and differences are less than 1\%.

In the case of a super-critical shock, the velocity of the upstream material is sufficiently high that the radiation from the shock has saturated and the maximum temperature in the precursor region now matches the temperature of the downstream (post-shock) region.  The precursor remains and, notwithstanding the leading edge, the gas and radiation are nearly in equilibrium.  Precursor and downstream regions are now connected via an embedded hydrodynamic shock and a ``Zel'Dovich spike'' \citep{mm84}, before conditions very quickly relax to the downstream state.

We set up a $M = 5$ super-critical shock tube in a similar manner to the sub-critical case.  A domain of $-4000 \leq x \leq 4000\, \mathrm{cm}$ is used, with the left state given by: $\rho_0 = 5.45887 \times 10^{-13}\, \mathrm{g\, cm^{-3}}$, $T_0 = 100\, \mathrm{K}$, $v_0 = 5.88588 \times 10^5\, \mathrm{cm\, s^{-1}}$, and the right state by: $\rho_1 = 1.96405 \times 10^{-12}\, \mathrm{g\, cm^{-3}}$, $T_1 = 855.720\, \mathrm{K}$, and $v_1 = 1.63592 \times 10^{5}\, \mathrm{cm\, s^{-1}}$.  As before, the initial discontinuity is shifted, in this case by $-310\, \mathrm{cm}$, to ensure the final shock position is $x = 0$.  The coarsest level has 256 zones, and 4 levels of refinement are used, giving an effective resolution of 4096 zones.  All other conditions are identical to the sub-critical test.

The right side of Figure \ref{fig:radshk} show our results at $t = 0.1\, \mathrm{s}$ for a super-critical $M = 5$ shock tube, plus the relative differences with respect to the semi-analytic solution.  Again, with the exception of the embedded shock region, the numerical and semi-analytic results are in good agreement.
\subsection{2-D: Protoplanetary disk relaxation benchmarks}
\label{ssub:diskrelax}
The primary science goal behind implementing FLD and irradiation in \azeus is to model protoplanetary disks and their evolution.  Towards this end, we now examine the structure of an idealised accretion disk which includes not only radiative cooling, viscous heating, and stellar irradiation, but also dynamics.  Inspired by \citet{kolbetal13}, we initialise a disk in axisymmetric (2.5-D) spherical $(r, \theta)$ coordinates, with computational domain $0.4 \leq r/a_\mathrm{Jup} \leq 2.5$ and $90 \leq \theta \leq 97^\circ$, where $a_\mathrm{Jup}$ is the semi-major axis of Jupiter.  To facilitate comparison with \citet{kolbetal13}, we use a uniform numerical resolution of $r\, \times\, \theta = 256\, \times\, 32$ zones (\ie, AMR is not employed).  We have, however, confirmed with higher resolution models that the results presented here are converged.

\begin{figure}[htb]
  \resizebox{\hsize}{!}{\includegraphics{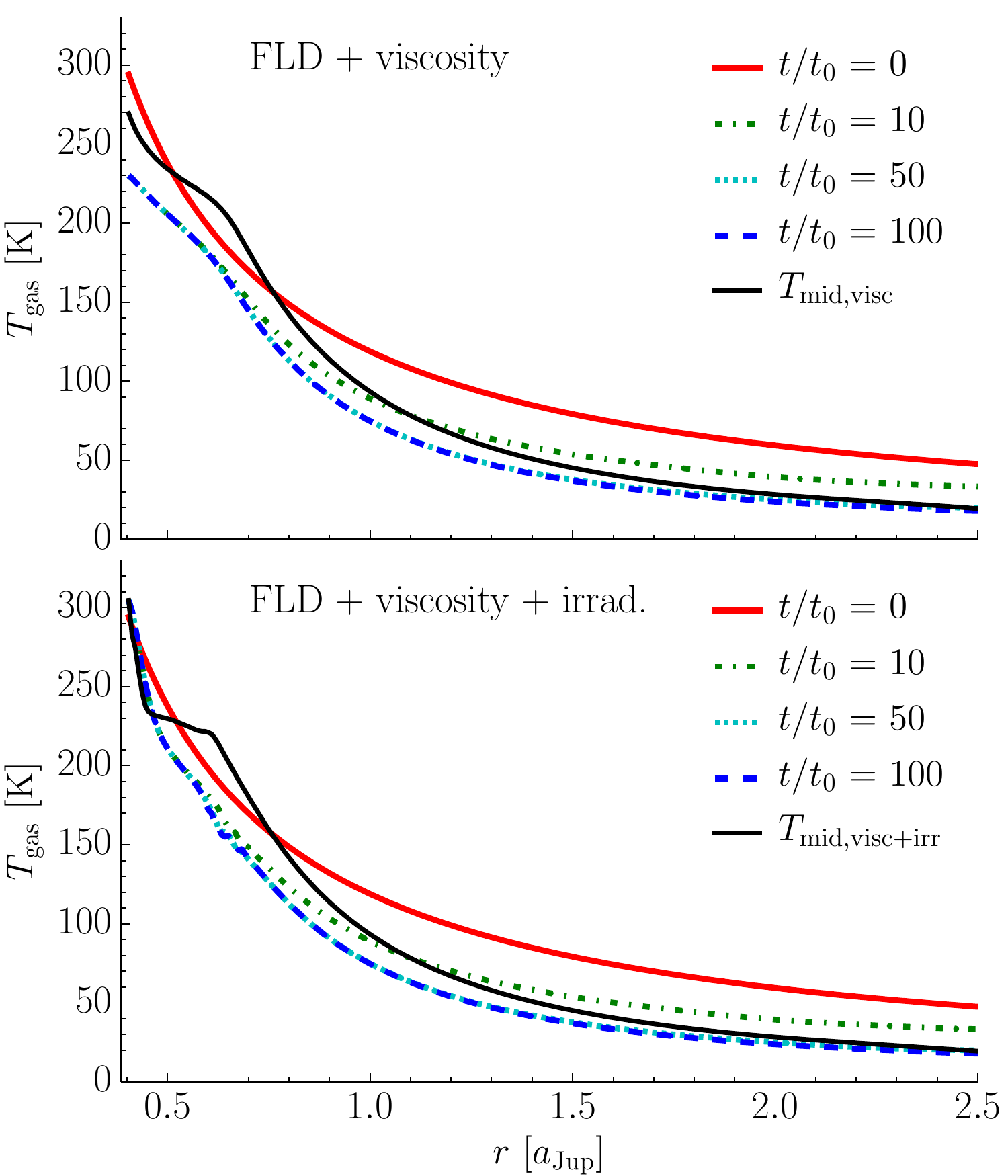}}
  \caption{Midplane gas temperatures in the PPD relaxation tests. \textit{Top:} the temperature as a function of radius at different times ($t_0 = 3.732 \times 10^8\, \mathrm{s}$) for the model including radiative cooling and viscous heating. \textit{Bottom:} the midplane temperature for the model which also includes stellar irradiation.  Temperatures predicted by Eqs.\ (\ref{eq:tmidvisc}; \textit{top}) and (\ref{eq:tcombined}; \textit{bottom}) are shown in black.}
  \label{fig:diskrelax}
\end{figure}

The initial density profile is given by:
\begin{equation}
  \label{eq:diskrho}
  \rho (r,\theta) = \rho_0(R) \exp \left(\frac{\sin \theta - 1}{h^2}\right),
\end{equation}
where $R = r \sin \theta$ is the cylindrical radius, $h = H / R = 0.05$ is the disk aspect ratio, $H$ is the disk scale height, $\rho_0(R) = \Sigma(R) / \sqrt{2\pi} H(R)$ is the midplane density, $\Sigma(R) = \Sigma_0 (R / a_\mathrm{Jup})^{-1/2}$ is the surface density, and $\Sigma_0$ is scaled such that the mass in the upper hemisphere of the disk is $0.005M_\odot$.

The initial pressure profile is given by $p = \rho c_\mathrm{iso}^2 = H\Omega_\mathrm{K} = H v_\mathrm{K} / R$, where $c_\mathrm{iso}$ is the isothermal sound speed, $\Omega_\mathrm{K}$ is the Keplerian orbital frequency, and $v_\mathrm{K} = \sqrt{GM_* / R}$ is the Keplerian speed.  The poloidal velocity is initially zero ($v_r = v_\theta = 0$), but the disk is rotationally supported by a slightly sub-Keplerian toroidal velocity:
\begin{equation}
  \label{eq:diskvphi}
  v_\phi = \sqrt{1 - 2h^2}v_\mathrm{K}.
\end{equation}
Pressure and centrifugal forces are offset by a point source gravitational potential with $M_* = 1M_\odot$, centred at $r = 0$.  The temperature is given by the ideal gas law, with $\gamma = 5/3$, $\mu = 2.3$.

Meanwhile, for the radiation, we use a Rosseland mean opacity given by the piece-wise functions of \citet{linpapaloizou85}; $\kappap = \kappar$ is assumed.  We also assume radiation and gas are initially in thermal equilibrium.  The flux limiter from \citet{kley1989}, which is optimised for disks, is applied.

With respect to boundary conditions, the toroidal velocity is set to the local Keplerian value at the inner and outer radial boundaries, and reflecting elsewhere.  All other hydrodynamic boundary conditions are set to reflecting.  For the radiation energy density, we set the upper $\theta$ boundary to $\er = \arad T_\mathrm{amb}^4$, with $T_\mathrm{amb} = 5\,\mathrm{K}$, and reflecting elsewhere.  This prescription allows the disk to radiatively cool unimpeded at the upper boundary.

Two different test cases of a radiative disk are considered here: first, a disk with internal viscous heating only, followed by a disk with both viscous and stellar irradiation heating.  In both cases, a constant kinematic viscosity of $\nu = 10^{15}\, \mathrm{cm^2\, s^{-1}}$ is applied \citep{kolbetal13}, and the system is evolved for 100 orbits at the radius of Jupiter (\ie, 1 orbit $= t_0 = 3.732 \times 10^8\, \mathrm{s}$).  

In the first case, an equilibrium state should be reached when radiative cooling, with a rate given by:
\begin{equation}
  \label{eq:radcool}
  Q^{-}_\mathrm{rad} = 2\sigma T^4_\mathrm{eff},
\end{equation}
balances the viscous heating, given by:
\begin{equation}
  \label{eq:vischeat}
  Q^{+}_\mathrm{visc} = \frac{9}{4} \Sigma \nu \Omega^2_\mathrm{K},
\end{equation}
where both rates are per unit area and take into account the two sides of the disk; $\sigma$ is the Stefan-Boltzmann constant.  Equating the two rates and solving for the effective temperature gives:
\begin{equation}
  \label{eq:diskteff}
  T_\mathrm{eff}^4(R) = \frac{9}{8} \frac{\nu\Sigma(R)\Omega_\mathrm{K}^2(R)}{\sigma},
\end{equation}
which describes the temperature of the gas at the disk surface.  To determine the temperature at the disk midplane, we first assume that all of the viscous heating occurs at the midplane, and then invoke radiative diffusion in the vertical direction to give (\eg, \citealt{armitage10book}):
\begin{equation}
  \label{eq:tdiffeq}
  \frac{\del T^4(z,R)}{\del z} = -\frac{3\rho(z,R)\kappar(z,R)}{8\sigma}Q^{+}_\mathrm{visc}.
\end{equation}
Integrating from the disk surface to the disk midplane, taking into account the variation in density and opacity (\ie, using \citealt{linpapaloizou85} opacities), the midplane temperature can be expressed as
\begin{align}
  \label{eq:tmidvisc}
  T^4_\mathrm{mid, visc}(R) =&~ T^4_\mathrm{eff}(R) \left(1 + \frac{3}{4}\int_0^\infty \rho(z,R)\kappar(z,R) dz \right)\nonumber\\
  =&~ T^4_\mathrm{eff}(R) \left(1 + \frac{3}{4}\tau_z(R) \right),
\end{align}
where equilibrium between viscous heating and radiative cooling has been exploited, and $\tau_z(R)$ is the vertical optical depth.

The top panel of Figure \ref{fig:diskrelax} plots the midplane temperature as a function of radius in a viscously heated disk at different times.  Also plotted (in black) is the midplane temperature predicted by Eq.\ (\ref{eq:tmidvisc}) for the surface density and vertical optical depth from the $t = 100t_0$ model.  As can be seen, by 50 orbits, the temperature has effectively reached a steady-state.  The numerical results, however, do not agree with the analytical predictions of Eq.\ (\ref{eq:tmidvisc}), with a factor of roughly $2^{1/4}$ difference.

\begin{figure*}[htb]
  \resizebox{\hsize}{!}{\includegraphics{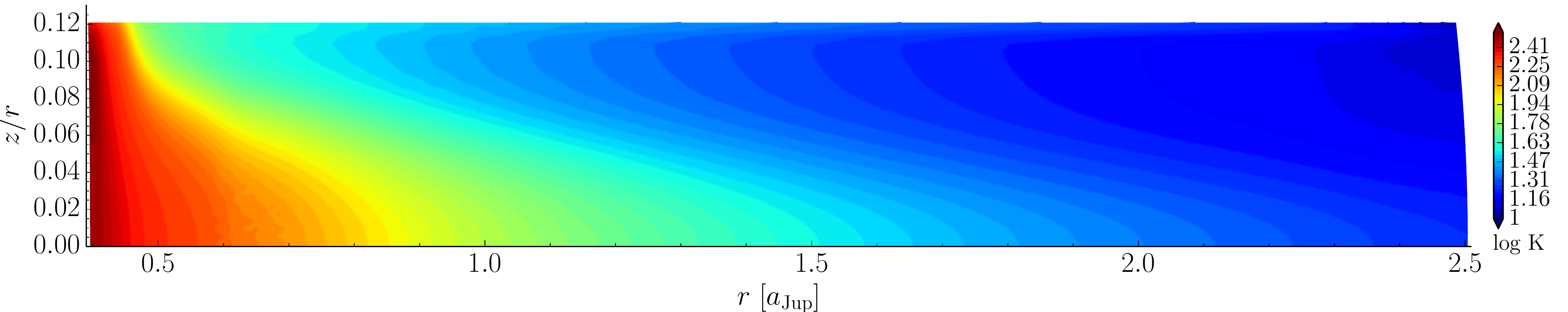}}
  \caption{2-D gas temperature structure at $t = 100t_0$ (filled colour contours) in the PPD relaxation test including radiative cooling, viscous heating, and stellar irradiation.}
  \label{fig:kolbdisktemp}
\end{figure*}

These differences can be attributed to two competing effects:  First, the assumption that all of the viscous heating occurs at the midplane is false, and this will tend to decrease the disk midplane temperature.  Second, in Eq.\ (\ref{eq:tdiffeq}), the Eddington approximation is implicitly assumed ($\lambda = 1/3$), even though we employ the flux limiter of \citet{kley1989} in these models.  Near the midplane, even though the Eddington approximation does roughly hold, it certainly does not in the upper layers of the disk (where \mbox{$\lambda \ll 1/3$}).  This limits the rate at which energy diffuses out of the disk, and will cause an increase in the midplane temperature.  Unfortunately, analytically accounting for both of these effects is not straightforward.  Yet, although the numerical results do not agree with the simple analytical model, they are in good agreement with previous numerical studies \citep{kolbetal13}.

In the second test case, heating by stellar irradiation is added to the model.  We consider only frequency-independent (grey) irradiation by a star with effective temperature $T_* = 5000\,\mathrm{K}$ and radius $R_* = 3R_\odot$.  The flux at the stellar surface is then given by $F_*(R_*) = \sigma T^4_*$.  Following \citet{bitschetal13}, the stellar opacity is set to $\kappa_* = 0.1 \kappap$.  We also assume that the disk continues inside the inner radial boundary by attenuating the stellar flux in the boundary zones using values of $\kappap$ and $\rho$ from the first active zone, effectively shielding the disk midplane from direct stellar heating (but not the upper layers).

With the addition of irradiation, the disk midplane temperature should reach a new equilibrium state, given by the combined balance of viscous heating, irradiation, and radiative cooling.  Under the assumption that the stellar irradiation is only absorbed at a single height in the disk, the temperature of this layer should be given by the equilibrium between stellar heating and radiative cooling.  The stellar heating rate per unit area is given by (\cf\ Eqs.\ \ref{eq:irrcases} and \ref{eq:fluxirr}):
\begin{align}
  \label{eq:qirrad}
  Q^{+}_\mathrm{irr} =&~ 2 \nabla\cdot\left(F_*(R_*) \frac{R_*^2}{r^2} e^{-\tau_*(r)}\right)\\
  =&~ 2 F_*(R_*) \frac{R_*^2}{r^2} e^{-\tau_*(r)} \left(1 - e^{-\Delta\tau_*(r)}\right),
\end{align}
where $\Delta\tau_*(r) = \rho(r)\kappa_*(r)\Delta r$, and accounting for both sides of the disk.  Equating this to the radiative cooling rate (Eq.\ \ref{eq:radcool}) then gives a temperature of
\begin{equation}
  \label{eq:tsurfirr}
  T^4_\mathrm{surf, irr}(r) = T_*^4 \frac{R_*^2}{r^2} e^{-\tau_*(r)} \left(1 - e^{-\Delta\tau_*(r)}\right).
\end{equation}

With the further assumption that, in the presence of irradiation heating only, the disk will be locally isothermal with a temperature profile given by Eq.\ (\ref{eq:tsurfirr}), then the midplane temperature of a disk including both viscous heating and irradiation can be estimated from Eqs.\ (\ref{eq:tmidvisc}) and (\ref{eq:tsurfirr}) as:
\begin{equation}
  \label{eq:tcombined}
  T^4_\mathrm{mid, visc+irr}(R) = T^4_\mathrm{mid, visc}(R) + T^4_\mathrm{surf, irr}(R)
\end{equation}

The bottom panel of Figure \ref{fig:diskrelax} shows the midplane temperature from the model including both viscous heating and stellar irradiation.  Also plotted (in black) is the midplane temperature predicted by Eq.\ (\ref{eq:tcombined}).

As in the purely viscous heating model, the temperature reaches a steady-state by $t = 50t_0$.  The disk is flat ($H_p / R \simeq$ constant; $H_p$ is the height where the majority of stellar irradiation is absorbed), and virtually all of the stellar irradiation is absorbed at low radii, leading to higher temperatures at low $R$ and a temperature profile in the outer disk that is dominated by viscous heating.  This is evidenced by similar midplane temperatures in the two panels of Figure \ref{fig:diskrelax} at large radii.  

While the numerical results generally agree with the temperatures predicted by Eq.\ (\ref{eq:tcombined}) at very low radii, they otherwise disagree.  Some of the disagreement is certainly due to the effects discussed above, but these cannot entirely explain the observed differences.  Figure \ref{fig:kolbdisktemp} shows the 2-D temperature structure of the disk model including stellar heating at $t = 100t_0$.  Clearly, the disk is not locally isothermal as assumed above.  Furthermore, Eq.\ (\ref{eq:tcombined}) does not consider any (radial or vertical) diffusion of the stellar heating.  Taking these in concert with the effects discussed for the first case, the differences between numerical and analytical results in Figure \ref{fig:diskrelax} are not surprising.
\subsection{2-D: static disk radiative transfer benchmarks}
\label{ssub:staticdiskrt}
\begin{figure*}
  \centering
  \resizebox{0.95\hsize}{!}{\includegraphics{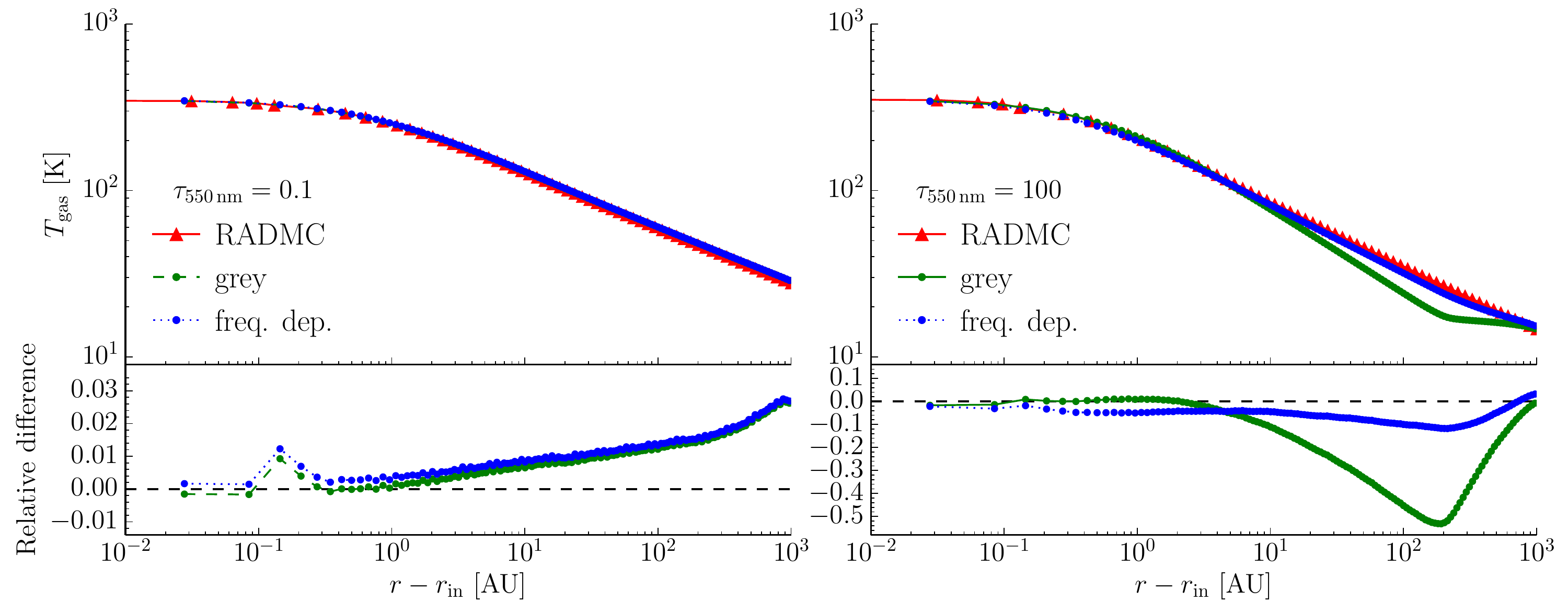}}
  \caption{Midplane gas temperatures in the low and moderate optical depth static disk RT benchmarks (following \citealt{pascuccietal04}). \textit{Top:} the temperature as a function of radius as determined by RADMC as well as both grey and frequency-dependent results as determined by \azeus.  \textit{Bottom:} the relative difference between \azeus and RADMC results.  Note that the RADMC and \azeus grid points are not co-spatial, and so the RADMC data is linearly interpolated to the location of \azeus zones as necessary.}
  \label{fig:pascuccitemps}
\end{figure*}

One of the motivations behind using a hybrid ray-tracing plus FLD algorithm is to significantly improve the results for stellar irradiated disks relative to pure FLD.  Herein, we present tests following \citet{kuiperetal10,kuiperklessen13} that serve as a comparison of the hybrid algorithm to more exact radiative transfer algorithms, as applied to protoplanetary disks.  More specifically, we simulate the static disks described in \citet{pascuccietal04,pinteetal09}, and compare the results to those produced with the RADMC Monte Carlo radiative transfer code \citep{dullemonddominik04}.

The first set of tests follows \citet{pascuccietal04}, where the density structure of a disk in cylindrical coordinates is given by:
\begin{equation}
  \label{eq:pascuccidisk}
  \rho(R,z) = \rho_0 \left(\frac{R}{R_0}\right)^{-1}\!\!\exp\left(-\frac{\pi}{4}\left(\frac{z}{h(R)}\right)^2\right),
\end{equation}
where $h(R) = z_0\, (R/R_0)^{1.125}$ is the scale height, $R_0 = 500\,\mathrm{AU}$, $z_0 = 125\,\mathrm{AU}$, and $\rho_0$ is used to scale the disk mass accordingly.  The temperature is assumed to be initially locally isothermal with profile $T(R) = T_0(R/1000\,\mathrm{AU})^{-1/3}$, where $T_0 = 14.7\,\mathrm{K}$, and is related to the gas energy by the ideal gas law (with $\gamma = 5/3$ and $\mu = 1$).  The diffuse radiation is also assumed to initially be in equilibrium with the gas, although this will not generally be the case for the converged results.

For the stellar irradiation, we use a central source with effective temperature $T_* = 5800\,\mathrm{K}$ and stellar radius $R_* = R_\odot$.  Opacities are given by \citet{drainelee84} for astronomical silicates, with a grain radius of $0.12\, \mu m$ and dust bulk density of $3.6\, \mathrm{g\, cm^{-3}}$.  The opacity table used has 61 frequency bins between $0.12$ and $2000\, \mu m$ \citep{pascuccietal04}.  A constant dust-to-gas ratio of $0.01$ is also assumed.  For the diffuse radiation component, the Planck and Rosseland mean opacities are calculated by integrating the dust opacities over frequency using the canonical definitions for each mean opacity.

We apply spherical coordinates for these tests, with a domain extending from $r_\mathrm{in} = 1 \leq r \leq 1000\,\mathrm{AU}$ and from $\pi /4 \leq \theta \leq 3\pi /4$, with 128 logarithmically spaced zones in the $r$-direction, and 180 uniformly spaced zones in the $\theta$-direction.  As the models are static, all physical modules other than irradiation and FLD are deactivated.  Therefore, the only boundary conditions we need are for the gas and radiation energies.  For the gas temperature, zero-gradient conditions are applied at all boundaries except for the outer radial boundary, where a value of $14.7$ K is maintained at all times.  The same is true for the radiation energy density, except that vacuum boundary conditions are applied at the inner radial boundary.  It is also important to note that, in contrast to \S \ref{ssub:diskrelax}, the inner edge of the disk is sharp and there is no material present inside $r_\mathrm{in}$.  Furthermore, scattering is neglected in both \azeus and RADMC models.

We calculate models using both FLD + frequency-dependent stellar irradiation (with frequency bins given by the opacity data) and FLD + grey stellar irradiation, where a stellar Planck mean opacity is calculated using the effective stellar temperature and opacity table.  In order to consistently obtain solutions for these tests, we use a direct LU method (rather than BiCGSTAB) to solve the radiation matrix\footnote{This illuminates one of the benefits of using the \petsc library to solve the radiation matrix, \ie, the ability to quickly and easily switch between different linear solvers.}.  Although slow, it has proven the most reliable method for these benchmarks, particularly at the extremely high optical depths of the \citet{pinteetal09} disks presented below.  As it is a direct solver, the parameter \texttt{slstol} does not apply.

\begin{table*}[htb]
  \centering
  \caption{Parameters and results for the static disk RT benchmarks.}
  \begin{tabular}{lllll}
  \hline\hline
  \multicolumn{3}{c}{\phantom{a}} & \multicolumn{2}{c}{Max.\ rel.\ diff.}\\
  $M_\mathrm{dust} (M_\odot)$ & $\rho_{0}$ (g cm$^{-3}$) & Midplane optical depth & grey & freq.-dep.\\
  \hline
  \rule{0pt}{2.0ex}
  \tablefootmark{a}$1.1 \times 10^{-7}$ & $8.107 \times 10^{-21}$ & $\tau_{550\,\mathrm{nm}}$ $= 0.1$ & 0.027 & 0.028\\
  \tablefootmark{a}$1.1 \times 10^{-4}$ & $8.107 \times 10^{-18}$ & \phantom{$\tau_{550\,\mathrm{nm}}$} $= 100$ & 0.53 & 0.12\\
  \tablefootmark{b}$3 \times 10^{-8}$ & $2.874 \times 10^{-19}$ & $\tau_{810\,\mathrm{nm}}$ $= 1.22 \times 10^{3}$ & 0.41 & 0.41 \\
  \tablefootmark{b}$3 \times 10^{-7}$ & $2.874 \times 10^{-18}$ & \phantom{$\tau_{810\,\mathrm{nm}}$} $= 1.22 \times 10^{4}$ & 0.45 & 0.45 \\
  \tablefootmark{b}$3 \times 10^{-5}$ & $2.874 \times 10^{-16}$ & \phantom{$\tau_{810\,\mathrm{nm}}$} $= 1.22 \times 10^{6}$ & 0.53 & 0.53 \\
  \hline
  \end{tabular}
  \tablefoot{References. \tablefoottext{a}{\citet[see also eq.\ \ref{eq:pascuccidisk}]{pascuccietal04}}\!, or \tablefoottext{b}{\citet[see also eq.\ \ref{eq:pintedisk}]{pinteetal09}}.  The parameter $\rho_0$ is determined from $\rho(R,z)$, $M_\mathrm{dust}$, and the dust-to-gas ratio.}
  \label{tab:diskdata}
\end{table*}

Figure \ref{fig:pascuccitemps} plots the results for disks of mass $M_\mathrm{dust} = 1.1 \times 10^{-7}$ and $1.1 \times 10^{-4} M_\odot$, which correspond to optical depths of $\tau_\mathrm{550\,nm} = 0.1$ and $100$, respectively (see Table \ref{tab:diskdata}, \citealt{pascuccietal04}).  To obtain these results, we iterate on the temperature structure until the relative change between successive calls to the radiation solver decreases below $10^{-4}$ (\cf\ \citealt{kuiperetal10}).

At low optical depths (Figure \ref{fig:pascuccitemps}; left), diffusion of radiation is unimportant, and the temperature structure of the disk is determined by stellar irradiation.  In this case, the \azeus results match very well with the corresponding RADMC models.  Furthermore, as the disk is optically thin in all frequency bins, grey and freq.-dep.\ models give very similar answers.  At moderate optical depths (Figure \ref{fig:pascuccitemps}; right), the differences between grey and freq.-dep.\ results are suddenly quite evident.  In this particular situation, the use of the Planck mean opacity in the grey model either over- or underestimates the stellar opacity at ultraviolet or infrared wavelengths, respectively, and as a result, the photon penetration depth varies significantly with frequency.  Clearly, the $\tau_{550\,\mathrm{nm}} = 100$ disk is as an excellent example of when frequency-dependent treatment of stellar irradiation matters.  Meanwhile, \citet{kuiperklessen13} have already demonstrated that using purely FLD for these tests gives results which are significantly worse than even the combination of grey irradiation and FLD.

The relative differences between \azeus and RADMC results are summarised in Table \ref{tab:diskdata}.  In addition, the \azeus results are generally in good agreement with \citet{kuiperklessen13}.  Differences are seen at large radii, but this may be due to the assumed initial temperature profile, which undergoes only minimal evolution in the outer disk.  

The second set of (higher optical depth) disk benchmarks performed have a density structure as described in \citet{pinteetal09}:
\begin{equation}
  \label{eq:pintedisk}
  \rho(R,z) = \rho_0 \left(\frac{R}{R_0}\right)^{-2.625}\!\!\!\!\!\!\!\exp\left(-\frac{1}{2}\left(\frac{z}{h(R)}\right)^2\right),
\end{equation}
where $r_0 = 100\,\mathrm{AU}$, $z_0 = 10\,\mathrm{AU}$, and $h(R)$ is as before.  Meanwhile, for the stellar component, an effective temperature of $T_* = 4000\,\mathrm{K}$ and a stellar radius of $R_* = 2 R_\odot$ are used.  The opacity table used is taken from \citet{weingartnerdraine01} for astronomical silicates of radius 1 $\mu\mathrm{m}$ and bulk grain densities of $3.5\,\mathrm{g\, cm^{-3}}$.  This opacity table contains 100 wavelength bins between $0.1$ and $3000\, \mu\mathrm{m}$.

\begin{figure*}
  \centering
  \resizebox{\hsize}{!}{\includegraphics{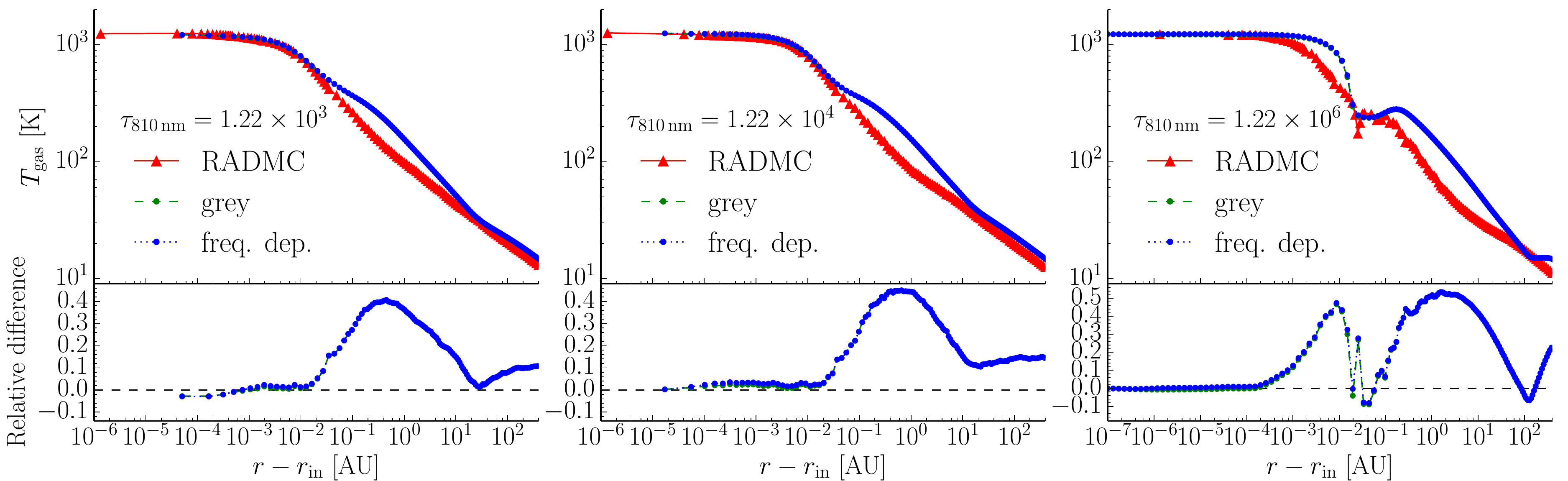}}
  \caption{Midplane gas temperatures in the high and extreme optical depth static disk RT benchmarks (following \citealt{pinteetal09}). \textit{Top:} the temperature as a function of radius as determined by RADMC as well as both grey and frequency-dependent results as determined by \azeus.  \textit{Bottom:} the relative difference between \azeus and RADMC results.}
  \label{fig:pintetemps}
\end{figure*}

The disks extend from $r_\mathrm{in} = 0.1 \leq r \leq 400\,\mathrm{AU}$, and $\pi /4 \leq \theta \leq 3\pi /4$, and have been scaled to midplane optical depths of $\tau_\mathrm{810\,nm} = 1.22 \times 10^3$, $1.22 \times 10^4$, and $1.22 \times 10^6$ (Table \ref{tab:diskdata}; \citealt{pinteetal09}).  To obtain the correct temperature structure, it is critical to numerically resolve strong gradients in optical depth, and in particular transitions from optically thin to thick.  Unfortunately, due to the extreme optical depths of the \citet{pinteetal09} disks, the simple grid set up used previously does not satisfy this criterion.  Therefore, we instead adopt the following set up:
\begin{description}
  \item[$\tau_{810\, \mathrm{nm}} = 1.22 \times 10^3$:] There are 20 ``ratioed'' zones in the $r$-direction from $0.1$ to $0.15$ AU, with a constant ratio of $\Delta r(i+1) / \Delta r(i) = 1.28$, resulting in a minimum zone size of $\Delta r = 1.01 \times 10^{-4}\,\mathrm{AU}$ at $r_\mathrm{in}$.  From $0.15$ to $400$ AU, 100 logarithmically spaced zones are employed.  In the $\theta$-direction, we continue to use 180 uniform zones between $\pi /4$ and $3\pi /4$.
  \item[$\tau_{810\,\mathrm{nm}} = 1.22 \times 10^4$:] 29 ratioed zones are placed in the $r$-direction between $0.1$ to $0.15$ AU with $\Delta r(i+1) / \Delta r(i) = 1.22$, followed by 100 logarithmically spaced zones.  The minimum zone size is $\Delta r = 3.5 \times 10^{-5}\,\mathrm{AU}$.  In the $\theta$-direction, there are 20 ratioed zones between $\pi/4$ and $3\pi /8$, followed by 130 uniform zones from $3\pi /8$ to $5\pi/8$, and subsequently another 20 ratioed zones.  The minimum $\Delta \theta$ in the ratioed zones is set to match smoothly with the uniform zones.
  \item[$\tau_{810\,\mathrm{nm}} = 1.22 \times 10^6$:] 60 ratioed zones from $0.1$ to $0.15$ AU, with $\Delta r(i+1) / \Delta r(i) = 1.31$ ($\Delta r_{\min} = 1.4 \times 10^{-9}\,\mathrm{AU}$), followed by 100 logarithmically spaced zones.  In the $\theta$-direction, we employ 50 ratioed zones between $\pi/4$ and $7\pi /10$, 500 uniform zones in the middle, followed by another 50 zones between $13\pi /20$ and $3\pi /4$.
\end{description}

We have chosen not to use AMR for these tests because the extreme resolution required at the inner edge of the disk clashes with how grids are refined in the AMR framework.  For example, in the $\tau_{810\,\mathrm{nm}} = 1.22 \times 10^{3}$ case, 9 levels of refinement with a refinement ratio of 4 would be required to obtain the same resolution at the inner edge, resulting in many more zones than needed using the above prescription.  The situation worsens considerably for even higher optical depths.

Figure \ref{fig:pintetemps} plots the results for the \citet{pinteetal09} benchmarks with masses of $M_\mathrm{dust} = 3 \times 10^{-8}$, $3 \times 10^{-7}$, and $3 \times 10^{-5}M_\odot$.  In these tests, the optical depths are sufficiently high that the disk is opaque to stellar photons at all frequencies, and the grey and freq.-dep.\ results become virtually identical.

For all the tests in this set, it can be seen that the \azeus results overestimate the midplane temperature between $(r - r_\mathrm{in}) \simeq 0.1$ -- $10\,\mathrm{AU}$ by up to $\sim\! 50\%$.  This region corresponds to where the grey vertical optical depth transitions through one, and thus we are seeing a similar effect as in the $\tau_{550\,\mathrm{nm}} = 100$ disk with regard to the importance of frequency-dependent optical depths.  In order to improve upon these results, we would need to adopt at least a frequency-dependent (\ie, multi-group) FLD approach.

The most extreme case, $\tau_{810\,\mathrm{nm}} = 1.22 \times 10^6$, is a particularly challenging problem for any radiative transfer code.  For example, \citet{pinteetal09} report that some of the codes included in the comparison did not converge for this particular test.  Indeed, from the right panel of Figure \ref{fig:pintetemps}, it can be seen that RADMC has difficulty converging to a solution, even after we increase the number of photon packets by a factor of 10 relative to the other tests.  For the \azeus results, we find that setting \texttt{nrtolrhs} $= 10^{-14}$ is necessary in order to consistently obtain convergence during the Newton-Raphson iterations.  Furthermore, it takes $> 2$ orders of magnitude more wall clock time to reach convergence than for the lower opacity tests.  While the right panel of Figure \ref{fig:pintetemps} presents the results when the relative change in temperature drops below $10^{-4}$ (as before), we have continued running the models and find that the ``dip'' in the midplane temperature between $0.1$ -- $10\,\mathrm{AU}$ disappears with $\sim\! 30\%$ more iterations.  That said, the \azeus results remain in reasonable agreement with the RADMC results, except (as before) for the vertical optical depth transition region, and are also in reasonably good agreement with \citet{kuiperklessen13}.
\subsection{2-D: Shadow test}
\label{ssub:shadow}
One of the continual criticisms against FLD is its inability to cast shadows (\eg, \citealt{hayesnorman03}).  By not solving a conservation equation for the radiative flux, and further by using Eq.\ (\ref{eq:fick}) as a closure relation, the radiative flux follows gradients in $\er$, and results in radiation propagating around corners.  Here, we present a test derived from \citet{hayesnorman03} and \citet{jiangetal12} to demonstrate how the hybrid FLD + ray-tracing algorithm can alleviate this weakness in certain contexts.  Additionally, in the second part of this test, we demonstrate that the hybrid algorithm is suitable for dynamical studies of photoevaporation, \ie, thermal winds driven by radiation.

Inspired by \citet{jiangetal12}, we consider a 2-D Cartesian domain of size $-0.5 \leq x \leq 0.5\,\mathrm{cm}$ and $-0.25 \leq y \leq 0.25\,\mathrm{cm}$.  An overdense ellipse is centred at the origin with a density profile prescribed by:
\begin{equation}
  \label{eq:shadowrho}
  \rho(x,y) = \rho_0 + \frac{\rho_1 - \rho_0}{1 + \exp(10(r-1))},
\end{equation}
where $r = (x/a)^2 + (y/b)^2 \leq 1$, $a = 0.01\,\mathrm{cm}$, $b = 0.06\,\mathrm{cm}$, and $\rho_1 = 10\rho_0$.  The ambient medium is initialised to a gas temperature of $T_0 = 290\,\mathrm{K}$ and a density $\rho_0 = 1\,\mathrm{g\,cm^{-3}}$.  The entire domain is initially in pressure equilibrium, and the diffuse radiation is assumed to be in equilibrium with the gas.  The gas opacity is given by:
\begin{equation}
  \label{eq:shadowopacity}
  \alpha = \kappa\rho = \alpha_0 \left(\frac{T}{T_0}\right)^{-3.5}\!\!\left(\frac{\rho}{\rho_0}\right)^2,
\end{equation}
where $\alpha_0 = 1$, and we have assumed that $\kappap = \kappar$.  We apply outflow boundary conditions for all variables and sides of the box, with the exception that we set a uniform and constant irradiation source across the left boundary (at $x = -0.5\,\mathrm{cm}$) characterised by a blackbody with effective temperature $T_\mathrm{irr} = 6T_0$.  Eq.\ (\ref{eq:shadowopacity}) is used for both the diffuse radiation and the irradiation.  For the grid, we employ 2 levels of refinement above the base grid, giving an effective resolution of $512 \times 256$ zones.  Refinement is based on gradients in the gas temperature and radiative energy density, both with thresholds of 5\%.

\begin{figure}[htb]
  \centering
  \resizebox{\hsize}{!}{\includegraphics{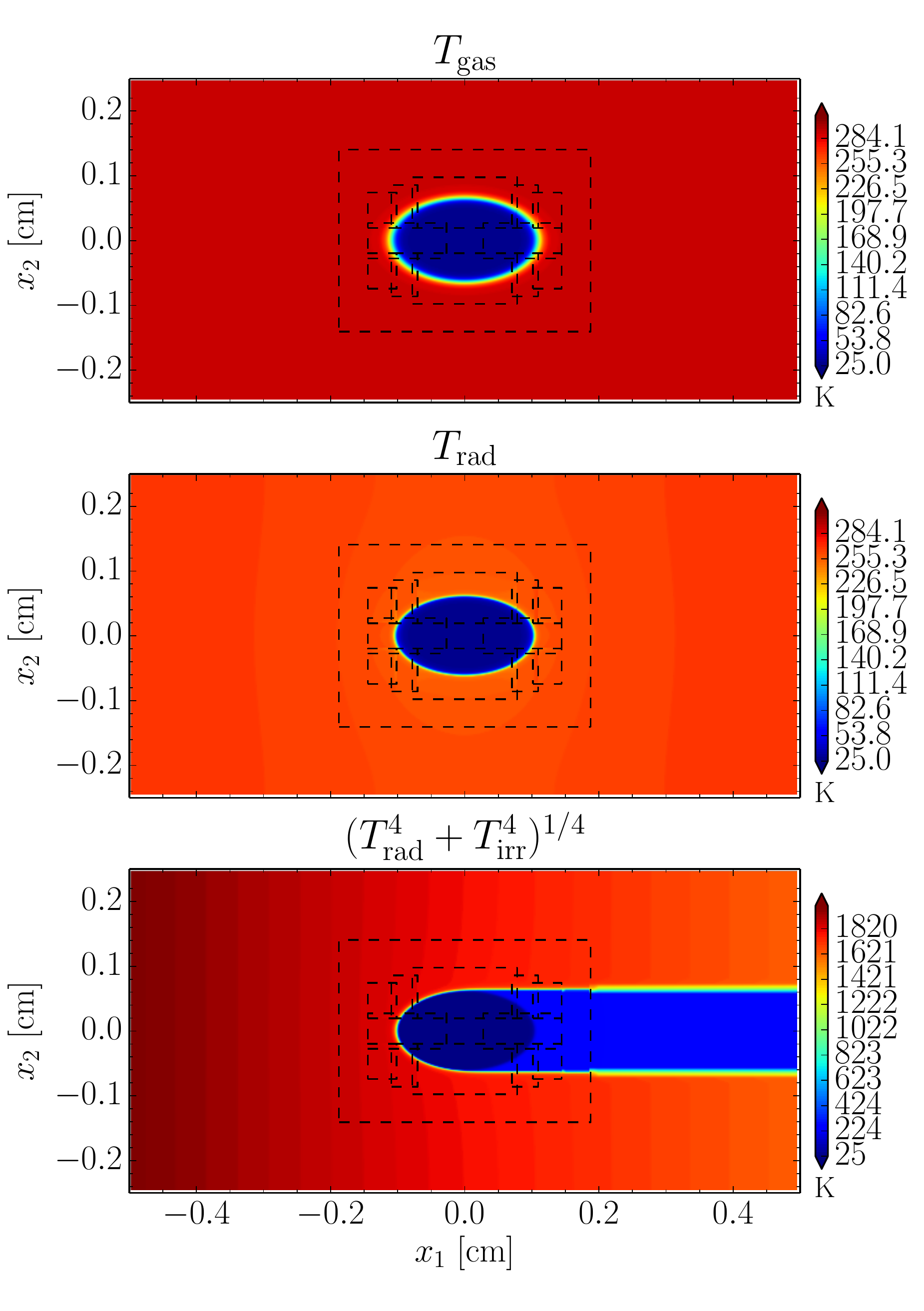}}
  \caption{Steady-state ($T_0 = 290\,\mathrm{K}$) shadow test results at $t = 3.3 \times 10^{-7}\,\mathrm{s}$, corresponding to $10^4$ light-crossing times in the ambient medium.  The source of irradiation is located at the left boundary.  \textit{Top:} gas temperature. \textit{Middle:} radiation temperature. \textit{Bottom:} ``total'' radiation temperature (diffuse + irradiation).  Dashed lines denote AMR grid boundaries.}
  \label{fig:shadowlow}
\end{figure}

In this set up, the ambient medium is optically thin, and the photon mean free path is equal to the length of the domain.  Meanwhile, the clump is optically thick, with an interior mean free path of $\simeq 3.2 \times 10^{-6}\,\mathrm{cm}$.  As a result, the clump will absorb the irradiation incident upon it, and should cast a shadow behind it.  Figure \ref{fig:shadowlow} shows the results using the hybrid FLD + ray-tracing algorithm at $t = 3.3 \times 10^{-7}\,\mathrm{s}$, or $10^4$ light-crossing times in the ambient medium.  Given these initial conditions, the dynamics are entirely negligible, and the solution quickly reaches a steady-state.  The bottom panel of Figure \ref{fig:shadowlow} shows the ``total'' radiation temperature ($= [T_\mathrm{rad}^4 + T_\mathrm{irr}^4]^{1/4}$, where $F_\mathrm{irr} = \sigma T_\mathrm{irr}^4$), and demonstrates not only the ability of the hybrid algorithm to cast shadows, but also the ability of the AMR to handle irradiation across changes in resolution.

\begin{figure*}[htb]
  \centering
  \resizebox{\hsize}{!}{\includegraphics[clip,trim=0 27 0 0]{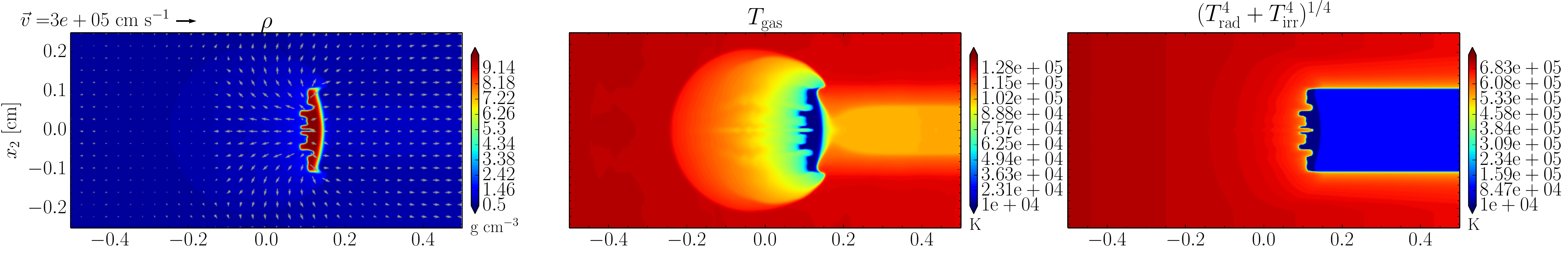}}
  \resizebox{\hsize}{!}{\includegraphics[clip,trim=0 0 0 27]{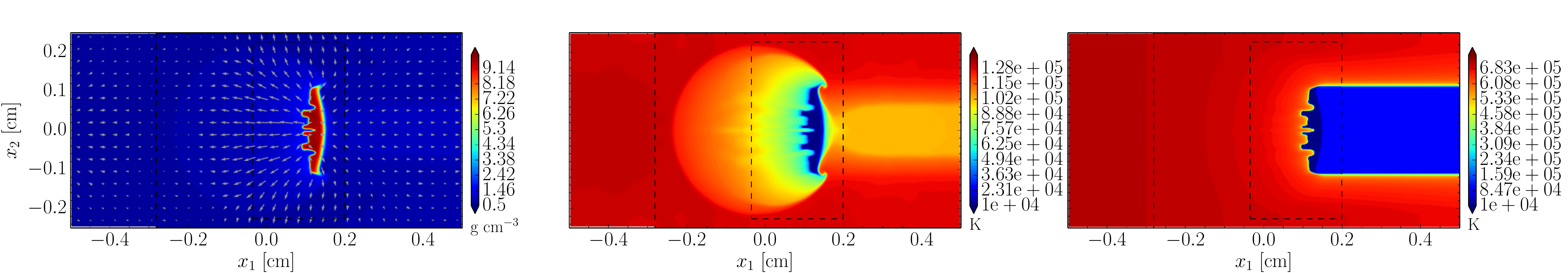}}
  \caption{Results for the evaporating shadow test ($T_0 = 103530\,\mathrm{K}$) at $t = 1.67 \times 10^{-6}\,\mathrm{s}$.  The irradiation source is located at the left boundary. \textit{Top:} Uniform grid results.  \textit{Bottom:} AMR results at the same time and effective resolution as the uniform grid.  \textit{Left:} gas density. \textit{Middle:} gas temperature. \textit{Right:} ``total'' radiation temperature (diffuse + irradiation).  Dashed lines denote AMR grid boundaries.}
  \label{fig:shadowevap}
\end{figure*}

To make things more interesting, and following \citet{jiangetal12}, we increase the initial ambient temperature by a factor of 357 ($\Rightarrow T_0 = 103530\,\mathrm{K}$) while keeping the remainder of the initial conditions the same.  In this case, heating due to irradiation (Eq.\ \ref{eq:irrcases} integrated over frequency) is no longer negligible, and heats the left side of the clump.  The irradiative heating establishes a pressure gradient and subsequent thermal wind which removes material from the clump over time (\ie, it photoevaporates).  Figure \ref{fig:shadowevap} shows the results of this process at $t = 1.67 \times 10^{-6}\,\mathrm{s}$ (5000 light-crossing times in the initial ambient medium) for both uniform and AMR grids, demonstrating that the AMR algorithms for irradiation and FLD are capable of reproducing the uniform solution.

Although the uniform and AMR results are qualitatively very similar, there are slight differences in the shape and position of the evaporation surface.  The shape and location of this surface is determined by the local irradiation flux integrated over many time steps, and the value of this flux is naturally sensitive to the column of material along the ray.  Any differences in the density along the ray between uniform and AMR models could then, over time, lead to a disagreement in the shape or position of the evaporation surface.  Densities are continually being interpolated in the AMR simulation to fill boundary zones for existing refined grids, or when a new refined grid is created, and these will not generally be equivalent to the densities at the same time and location in the uniform grid model.  It is therefore not surprising to find differences between uniform and AMR models, but it is reassuring that they remain small.

Finally, a cautionary statement:  In order to produce the results seen in the bottom panel of Figure \ref{fig:shadowevap}, we increased the number of AMR buffer zones (\texttt{ibuff}) while decreasing the grid efficiency parameter (\texttt{geffcy}) relative to the steady-state case (see Table \ref{tab:testparameters}).  For a test problem such as this, where the radiation mean free path is large in much of the domain, a nearly global solution at any given location may be necessary in order to obtain the correct solution.  As realised by other authors (\eg, \citealt{commerconetal14}), this can be problematic for any AMR-FLD algorithm which calculates a local solution and later applies corrections, such as the deferred synchronisation algorithm used here.  To minimise this issue, we have intentionally relaxed the values for \texttt{geffcy} and \texttt{ibuff} to favour a few large grids over many small grids.  The alternative would be to instead use a more complex, expensive and global, multi-level solver (\eg, \citealt{howellgreenough03}).

\section{Summary \& discussion}
\label{sec:discuss}
We have presented a new implementation for RHD in the \azeus AMR-MHD fluid code, combining two-temperature (non-equilibrium) FLD and frequency-dependent stellar irradiation using the simple and fast ray-tracing algorithm of \citet{kuiperetal10}.  The radiation subsystem is operator split from the hydrodynamics, and is solved implicitly using a globally convergent Newton-Raphson method.  Meanwhile, we employ the flexible and freely-available \petsc library \citep{petsc-web-page} to solve the radiation matrix inside the Newton-Raphson iterations.  Both FLD (via the deferred synchronisation algorithm of \citealt{zhangetal11}) and stellar irradiation are available for use with AMR and in curvilinear coordinates.  


The RHD implementation in \azeus inherits the general limitations of the FLD approximation, including difficulties with transitions from optically thick to thin, and the loss of directional information for the radiative flux.  However, in certain contexts, the combination with ray-tracing mitigates these shortcomings; the resultant hybrid algorithm is even capable of casting shadows.  Furthermore, FLD is typically much less computationally expensive than other, more accurate, radiative transfer methods (\eg, VET, \citealt{jiangetal12}; M1 moment method, \citealt{gonzalezetal07}; Monte Carlo methods, \citealt{haworthharries12}).

We have presented several benchmarks to validate our methods and demonstrated the usefulness of the code not only for RHD simulations of PPDs, but also more general astrophysical contexts.  These tests also demonstrate that \azeus is competitive with other available AMR-RHD grid codes, even for very challenging disk radiative transfer problems.  For additional information on the code, including results from additional numerical tests, we direct the reader to the code website: \url{http://www.ica.smu.ca/azeus/}.

The driving motivation behind implementing RHD in \azeus is to study photoevaporation in PPDs.  The combination of FLD and frequency-dependent stellar irradiation presented here provides us with the necessary foundations to do so.  In an upcoming paper (Ramsey, Bruderer, \& Dullemond, in prep.), we expand the code further to include a two-fluid approximation to enable the decoupling of dust and gas temperatures, as is expected in the upper layers of PPDs (\eg, \citealt{kampdullemond04}).  We also implement a simplified chemistry module, providing us with all the tools we need to examine far-UV powered (\ie, at energies $6 < h\nu < 13.6$ eV) photoevaporation in a time-dependent and self-consistent manner.

\begin{acknowledgements}
This work is supported by DFG grant DU 414/9-1.  We would like to thank the referee for a timely and constructive report.  JPR would like to thank Rolf Kuiper for helpful technical discussions on irradiation and the tests of \S \ref{ssub:staticdiskrt} during the early stages of this research, as well as Bertram Bitsch for useful discussions on what developed into the tests of \S \ref{ssub:diskrelax}. JPR would also like to acknowledge the NORDITA program on Photo-Evaporation in Astrophysical Systems (June 2013) where part of this work was carried out.
\end{acknowledgements}
\bibliographystyle{bibtex/aa}
\bibliography{./rdI.bib}
\end{document}